%% file: mem5pap.tex
\documentclass[aps,showpacs]{revtex4}
\usepackage{epsfig}
\usepackage{longtable}
\newcommand{\nc}{\newcommand}
\nc{\rnc}{\renewcommand}

\nc{\beq}{\begin{equation}}
\nc{\eeq}{\end{equation}}
\nc{\bea}{\begin{eqnarray}}
\nc{\eea}{\end{eqnarray}}
\nc{\bpi}{\begin{picture}}
\nc{\epi}{\end{picture}}
\nc{\ba}{\begin{array}}
\nc{\ea}{\end{array}}
\nc{\nn}{\nonumber}
\nc{\ts}{\textstyle}
\nc{\ds}{\displaystyle}
\nc{\bm}{\boldmath}

\nc{\p}{\partial}
\nc{\f}[2]{\frac{#1}{#2}}
\nc{\od}{{\cal O}}
\nc{\D}{{\cal D}}
\nc{\vx}{\mbox{\bm$x$}}
\nc{\gh}{\hat{g}}
\nc{\pp}{\phantom{+}}

\nc{\al}{\alpha}
\nc{\be}{\beta}
\nc{\ga}{\gamma}
\nc{\de}{\delta}
\nc{\ep}{\epsilon}
\nc{\ve}{\varepsilon}
\nc{\ka}{\kappa}
\nc{\om}{\omega}
\rnc{\th}{\theta}
\nc{\vp}{\varphi}
\nc{\Ga}{\Gamma}
\nc{\De}{\Delta}
\nc{\La}{\Lambda}

\begin{document}

\bibliographystyle{apsrev}

\title{Fluctuation Pressure of a Membrane Between Walls Through Five Loops}

\author{Boris Kastening}
\affiliation{Institut f\"ur Theoretische Physik\\
Freie Universit\"at Berlin\\ Arnimallee 14\\ D-14195 Berlin\\ Germany\\
email: {\tt ka@physik.fu-berlin.de}}

\date{3 May 2002}

\begin{abstract}
An earlier four-loop calculation of the fluctuation pressure of a fluid
membrane between two infinite walls is extended to five loops.
Variational perturbation theory is used to extract the hard-wall limit
from perturbative results obtained with a smooth potential.
Comparison with a structurally similar quantum mechanics problem of
a particle in a box is used for an alternative way of extracting the
membrane pressure and also to estimate the quality of the results.
Our values lie above the best available Monte Carlo data.
\end{abstract}

\maketitle

\section{Introduction}
The dominant repulsive force between layered chemical and biological
systems, called membranes, is given by thermal out-of-plane fluctuations
\cite{HeSe,BiPeSoOs}.
In the absence of tension, these membranes are called fluid and the
fluctuations are controlled by the membranes' bending rigidity.
There have been various theoretical approaches to compute the pressure
of a single membrane between walls \cite{HeSe,JaKl1,JaKlMe,GoKr,Kl,BaKlPe1}
or of a stack of membranes \cite{HeSe,JaKl1,JaKlMe,GoKr,onlystack}.
These situations are also interesting statistical mechanics problems.

Here we are concerned with the pressure generated by the bending
fluctuations of a fluid membrane between two infinitely extended parallel
walls, which has the form \cite{HeSe}
\beq
\label{pressure}
p=\al\f{(k_BT)^2}{\ka(d/2)^3},
\eeq
where $\ka$ is the bending rigidity of the membrane, $d$ is the distance
between the walls and $\al$ is a factor that we wish to compute.
Estimates of $\al$ have been ranging over the years from theoretical
approximations $\al\approx0.0242$ by Helfrich \cite{HeSe} and
$\al\approx0.0625$ by Janke and Kleinert \cite{JaKl1} through Monte Carlo
estimates $\al=0.079\pm0.002$ by Janke, Kleinert and Meinhart \cite{JaKlMe}
and $\al=0.0798\pm0.0003$ by Gompper and Kroll \cite{GoKr} and
a recent theoretical estimate $\al\approx0.0797$ by Bachmann,
Kleinert and Pelster \cite{BaKlPe1}.

In \cite{BaKlPe1}, the result was obtained by replacing the hard walls
by a $\tan^2$ potential, whose prefactor was sent to zero at the end
of the calculation to recover hard walls.
This corresponds to the strong-coupling limit
$\al=\lim_{g\rightarrow\infty}\al(g)$ of a loop expansion of $\al(g)$,
where $1/g^2$ is proportional to the prefactor of the potential.
To achieve the necessary resummation, variational perturbation theory
(VPT) \cite{pibook} was used.
This technique has been successful also in other situations where
the strong-coupling limit of an asymptotic weak-coupling series is
sought, e.g.\ when computing critical exponents from $\phi^4$
field theory models \cite{phi4book}.
In this work, we extend the four-loop calculation of \cite{BaKlPe1}
to five loops.

Our work is structured as follows.
In sec.\ \ref{bcmodel}, we model the hard walls with two different
potentials and give the perturbative results in sec.\ \ref{pt}.
In sec.\ \ref{vpt}, we follow \cite{BaKlPe1} and use VPT to estimate the
strong-coupling limit corresponding to hard walls.
In sec.\ \ref{QM}, we consider the two different potentials to model the
walls for a quantum mechanical (QM) particle in a box.
The problem of finding the ground state energy for the QM problem is
identical to computing the partition sum of a string between walls modeled
by the same potential \cite{Kl,dGLi,KlChHa}.
This problem in turn is structurally equivalent to finding $\al(g)$ in the
membrane problem.
Only for one of these potentials, the solution is known exactly
\cite{Kl,KlChHa}.
Although the potentials are very similar in the region of interest,
their behavior under resummation with VPT is rather different.
This will be used to judge the quality of the results of using VPT for
the membrane problem. 
In sec.\ \ref{memqm} we force $\al(g)$ for the membrane problem to
be identical to $\al(g)$ in the solvable QM problem by choosing the
potential appropriately and extract $\al$ from determining the potential's
singularities.
In sec.\ \ref{conclusions} we discuss our results.

\section{Modeling of the Boundary Conditions}
\label{bcmodel}
Consider a tensionless membrane between two large flat parallel walls of
area $A$ separated by a distance $d$.
In the harmonic approximation, which we are considering throughout, the
curvature energy is given by
\beq
\label{energy}
E=\f{\ka}{2}\int_A d^2x[\p^2\vp(\vx)]^2,
\eeq
where $\ka$ is the membrane's bending rigidity and $\vp$ is a field that
describes the membrane's position between the walls, which are located
at $\pm d/2$.
The $d$-dependent part $f_d$ of the free energy density of the system at
temperature $T$ is given by the path integral
\beq
\label{f}
\exp\left(-\f{A f_d}{k_BT}\right)
=\prod_{\vx}\int_{-d/2}^{+d/2}d\vp(\vx)\exp\left(-\f{E}{k_BT}\right).
\eeq
The pressure is then obtained as
\beq
p=-\f{\p f_d}{\p d}
\eeq
and has the form (\ref{pressure}) \cite{HeSe,JaKl1} and our task is to find
the constant $\al$.

The difficulty in computing the path integral (\ref{f}) consists in
implementing the restriction $-d/2<\vp<d/2$.
We follow \cite{Kl,BaKlPe1} and add a potential term
$m^4d^2\int d^2xV(\vp/d)$ to $E$, where $V$ has a sufficiently strong
singularity at $\pm1/2$, expand the potential $V$ in a Taylor series
in $\vp$ and drop the restriction on $\vp$.
At the end of the calculation we take the limit $m\rightarrow0$.
We consider the potentials
\beq
V_c(z)=\f{1}{2\pi^2\cos^2(\pi z)}
\eeq
and
\beq
V_a(z)=\f{1}{48}\left[\f{1}{(1+2z)^2}+\f{1}{(1-2z)^2}\right].
\eeq
The potentials have in common that they have quadratic divergences
at $\pm1/2$ and that their quadratic term in a Taylor expansion is
normalized to $z^2/2$.
$V_c$ is related to the potential $V_t=(2\pi^2)^{-1}\tan^2(\pi x)$ used
in \cite{BaKlPe1} by $V_c=(2\pi^2)^{-1}+V_t$.
For the resummation procedure employed in \cite{BaKlPe1}, $V_c$ and $V_t$
yield identical results and we will therefore recover the four-loop result
reported there.
For the other procedures used here, $V_c$ is better suited than $V_t$.

Since the functional form of $p$ in terms of $\ka$, $d$ and $T$ is known
and since we are going to differentiate only with respect to $d$, we will
set $k_BT=\ka=1$ in the sequel. 
The energy functional may then be expanded as
\beq
\label{epenergy}
E=\int d^2x\left\{\f{1}{2}[\p^2\vp(\vx)]^2+\f{1}{2}m^4\vp(\vx)^2
+m^4\ep_0d^2
+m^4\sum_{k=2}^\infty\ep_{2k}d^{2(1-k)}\vp(\vx)^{2k}\right\},
\eeq
where the $\ep_{2k}$ are the expansion coefficients of the potential.

The path integral can now be evaluated in a loop expansion \cite{Kl,BaKlPe1}.
The resulting Feynman diagrams, including their combinatorial factors, are
obtained from recursion relations, whose derivation is delegated
to appendix~\ref{loopexp}.
The evaluation of the associated momentum integrals is detailed in appendix
\ref{mbintegrals}.

\section{Perturbation Theory}
\label{pt}
The diagrams labeled $L{-}n$ ($n$th $L$-loop diagram) of
appendix \ref{loopexp} correspond apart from combinatorial factors $c_{L-n}$
and coupling constant factors $g_{L-n}$ to integrals in $\vx$-space.
Since the diagrams are connected and because of translational symmetry
in the infinite-wall limit, we may split off a factor $A$ from each
diagram and represent the remainder in momentum space.
A line represents then a propagator
\beq
\label{prop}
\De(p^2,m^2)=\f{1}{p^4+m^4}=\f{i}{2m^2}
\left(\f{1}{p^2+im^2}-\f{1}{p^2-im^2}\right),
\eeq
while the integration measure over all independent momenta is
\beq
\int_p\equiv\int\f{d^2p}{(2\pi)^2}
\eeq
with momentum conservation at each vertex.
A vertex with $2k$ legs represents a factor
\beq
\label{ccfactor}
-m^4d^{2(1-k)}\ep_{2k}.
\eeq
The sum of all diagrams corresponds to the negative of the free energy
density $f_m$, where the index refers to the presence of a
non-zero $m$ and $\lim_{m\rightarrow0}f_m=f_d$.

In the sequel, a diagram represents only the corresponding momentum space
integral which we call $I_{L-n}$, i.e.\ we split off not only a factor $A$,
but also the combinatorial factor $c_{L-n}$ and the $-\ep_{2k}$-part of the
factors (\ref{ccfactor}), which we collect into $g_{L-n}$.
Then $f_m$ has $L$-loop expansions
\beq
f_m=\f{1}{d^2}\sum_{l=0}^L a_lg^{l-2},
\eeq
with
\beq
a_L=-\sum_ng_{L-n}c_{L-n}I_{L-n}
\eeq
and a coupling constant
\beq
\label{gdefmem}
g=\f{1}{m^2d^2}.
\eeq
In Table \ref{diagrams}, we give $g_{L-n}$, $c_{L-n}$ and $I_{L-n}$
through five loops.
For instance, the resulting zero-, one- and two-loop contributions are
\beq
a_0=\ep_0,~~~~a_1=\f{1}{8},~~~~a_2=\f{3}{64}\ep_4.
\eeq
Through five loops, we get for the potentials under consideration
\beq
\label{memexpansion}
\begin{tabular}{l||l|l}
$L$ & \multicolumn{1}{c|}{$a_L$ for $V_c$}
& \multicolumn{1}{c}{$a_L$ for $V_a$}
\\\hline
0 & $\pp 0.0506606 $ & $\pp 0.0416667 $ \\
1 & $\pp 0.125000  $ & $\pp 0.125000  $ \\
2 & $\pp 0.154213  $ & $\pp 0.156250  $ \\
3 & $\pp 0.105998  $ & $\pp 0.102307  $ \\
4 & $\pp 0.026569  $ & $\pp 0.028101  $ \\
5 & $   -0.034229  $ & $   -0.031426  $ \\
\end{tabular}
\eeq

\section{\bm$\al$ From Strong-Coupling Variational Perturbation Theory}
\label{vpt}
The $d$-dependent part of the free energy has for $m^2=0$ the form
$f=4\al/d^2$, where the factor $4$ ensures consistency with
(\ref{pressure}).
Our task is to find an approximation to the strong-coupling limit
$\al=\lim_{g\rightarrow\infty}\al(g)$ with $L$-loop expansions of $\al(g)$
given by
\beq
\label{alw}
\al(g)=\f{1}{4g^2}\sum_{l=0}^L a_lg^l,
\eeq
with the knowledge of only the first few $a_l$.
We will assume that $\al(g)$ has a strong-coupling expansion
\beq
\label{als}
\al(g)=\f{1}{4}\sum_{m=0}^\infty a_m'g^{-2m/q}
\eeq
with an additional parameter $q$.
Then the problem has the following form:
Given a function $f(g)=4\al(g)$ with $L$-loop weak-coupling expansions
\beq
\label{weakexpansion}
f_L(g)=g^r\sum_{l=0}^Lf^w_lg^l
\eeq
and assuming strong-coupling expansions
\beq
\label{strongexpansion}
f^M(g)=g^{p/q}\sum_{m=0}^Mf^s_mg^{-2m/q},
\eeq
we are interested in finding $f_0^s$, $p$ and $q$.
Assuming a thermodynamic limit for the problem at hand means setting
$p=0$.
Then $\al$ exists and is given by $\al=f_0^s/4$.
In \cite{BaKlPe1} it was additionally assumed that $q=1$, which is
motivated by a similar QM problem, see sec.\ \ref{QM}.

In VPT \cite{pibook}, we replace in
(\ref{weakexpansion})
\bea
g^{l+r}
&\rightarrow&
(tg)^{l+r}\left\{\left(\f{g}{\gh}\right)^{2/q}
+t\left[1-\left(\f{g}{\gh}\right)^{2/q}\right]\right\}^{[p-(l+r)q]/2}
\nn\\
&=&
\left(\f{g}{\gh}\right)^{p/q}(t\gh)^{l+r}
\left\{1+t\left[\left(\f{\gh}{g}\right)^{2/q}-1\right]\right\}^{[p-(l+r)q]/2},
\eea
reexpand the resulting expression in $t$ through $t^{L+r}$, set $t=1$
and then optimize the resulting expression in $\gh$, where optimizing
refers to the principle of minimal sensitivity \cite{St} and in
practice means finding appropriate stationary or turning points.
That is, we replace
\bea
\label{gghat}
g^{l+r}
&\rightarrow&
\left(\f{g}{\gh}\right)^{p/q}\gh^{l+r}\sum_{k=0}^{L-l}
\left(\ba{c}[p-(l+r)q]/2\\k\ea\right)
\left[\left(\f{\gh}{g}\right)^{2/q}-1\right]^k
\eea
and optimize the resulting expression in $\gh$.
For $V_c$ and $V_a$, where $r=-2$, we obtain with $p=0$
\beq
\al_L=\f{1}{4}{\rm opt}_{\gh}\left[\sum_{l=0}^La_l\gh^{l-2}
\sum_{k=0}^{L-l}\left(\ba{c}(2-l)q/2\\k\ea\right)(-1)^k\right]
\eeq
as the $L$-loop variational approximation to $\al$.
This expression also holds for $V_t$, where $r=-1$.
For $q=1$, the expression in square brackets is independent of $a_0$,
which is why we reproduce below the results of \cite{BaKlPe1}.

If we do not want to make assumptions about $q$ for $f_m$, we can
determine it self-consistently by first treating $d\ln f_m^2/d\ln g$
in VPT, since it has the same $q$ as $f_m$ and since
\beq
\lim_{g\rightarrow\infty}\f{d\ln f_m^2}{d\ln g}=\f{p}{q}
\eeq
with $p=0$ by assumption of a thermodynamic limit.
That is, we resum the expansion of $d\ln f_m^2/d\ln g$ as detailed above
and tune $q$ such that optimization with respect to $\gh$ leads to 
$d\ln f_m^2/d\ln g=0$.

A similar QM problem (see sec.\ \ref{QM} below) leads us to try $q=1$.
Let us consider the different potentials for modeling the walls that
enclose the membrane with this assumption.
The loop orders $0$--$2$ do not admit a variational solution and
we therefore take the perturbative results as our best approximation.
Then the loop orders $0$ and $1$ yield zero for $\al$, since they
contain only negative powers of $g$, and the two-loop result is
$\al_2=a_2/4$.
The results through five loops are
\beq
\label{alvp}
\begin{tabular}{l||l|l}
$L$ & \multicolumn{1}{c|}{$\al$ for $V_c$}
& \multicolumn{1}{c}{$\al$ for $V_a$} \\
\hline
$2$ & 0.038553 & 0.039063 \\
$3$ & 0.073797 & 0.073688 \\
$4$ & 0.079473 & 0.079422 \\
$5$ & 0.081354 & 0.081345 \\
\end{tabular}
\eeq
with the results for $V_c$ through four loops coinciding with those reported
in \cite{BaKlPe1}.
An extrapolation of the results (\ref{alvp}) suggests a value of $\al$
between $0.0820$ and $0.0825$.

The results from determining $q$ self-consistently as described above are
through five loops
\beq
\begin{tabular}{c||l|l||l|l}
& \multicolumn{2}{c||}{$V_c$} & \multicolumn{2}{c}{$V_a$}
\\\hline
$L$
& \multicolumn{1}{c|}{$q$} & \multicolumn{1}{c||}{$\al$} 
& \multicolumn{1}{c|}{$q$} & \multicolumn{1}{c}{$\al$} \\
\hline
$3$ & $0.38124$ & $0.093076$ & & \\
$4$ & $0.56789$ & $0.095830$ & $0.46463$ & $0.098222$ \\
$5$ & $0.73907$ & $0.090983$ & $0.74209$ & $0.090321$ \\
\end{tabular}
\eeq
The values are compatible with convergence towards $q=1$ and with the
$\al$-values for $q=1$, but convergence is too slow for any quantitative
use.

\section{Quantum Mechanical Particle in a Box}
\label{QM}
A one-dimensional problem similar to the two-dimensional case above
is finding the ground state energy of a QM particle in a one-dimensional
box \cite{KlChHa}.
The Euclidean path integral to be computed is
\beq
\label{zqm}
\exp(-TE^{(0)})=\prod_{t}\int_{-d/2}^{+d/2}d\vp(t)\exp\left(-E\right)
\eeq
with
\beq
\label{stringenergy}
E=\f{1}{2}\int_{-T/2}^{+T/2}dt\dot{\vp}(t)^2,
\eeq
where $T$ is the total interaction time, being equivalent to the area $A$
in the membrane case.
In the large-$T$ limit, $E^{(0)}$ is the ground state energy, for which we
will test our approximation methods.
Its exact value is
\beq
E^{(0)}=\f{\pi^2}{2d^2}.
\eeq
Again, we model the walls with a potential,
\beq
E=\int_{-T/2}^{+T/2}dt\left[\f{1}{2}\dot{\vp}(t)^2+m^2d^2V(\vp(t)/d)\right],
\eeq
and as in the membrane case for $f_m/(k_BT)$, we can write down a loop
expansion for $E^{(0)}$.
After modifying the Feynman rules according to
\beq
\f{1}{p^4+m^4}\rightarrow\f{1}{p^2+m^2},~~~~~~
\int\f{d^2p}{(2\pi)^2}
\rightarrow\int_{-\infty}^{+\infty}\f{dp}{2\pi},~~~~~~
-m^4d^{2(1-k)}\ep_{2k}\rightarrow-m^2d^{2(1-k)}\ep_{2k}
\eeq
and defining $\al(g)$ and $g$ for the QM problem by
\beq
E^{(0)}=\f{64\al(g)}{d^2},~~~~~~g=\f{4}{md^2},
\eeq
not only can $\al(g)$ be expanded as in (\ref{alw}), but due to the simple
relation (\ref{qmmbsim}) between one-loop integrals in the membrane and
QM cases, all diagrams that separate into one-loop integrals give the same
contribution to $\al(g)$ in both cases \cite{Kl}.
It follows that for any given potential, $a_0$, $a_1$ and $a_2$, which
involve at most one-loop topologies, are identical in the QM and the
membrane problem.

For $V_c$, the exact ground state energy is known for any $m$ and $d$
\cite{KlChHa},
\bea
\label{ec}
E^{(0)}_c
&=&
\f{\pi^2}{2d^2}\left(\f{16}{\pi^4g^2}+\f{1}{2}
+\f{4}{\pi^2g}\sqrt{1+\f{\pi^4g^2}{64}}\right)
=\f{\pi^2}{2d^2}\left(\f{1}{2}+\f{16}{\pi^4g^2}
+\f{1}{2}\sqrt{1+\f{64}{\pi^4g^2}}\right).
~~~~
\eea
The limiting value for $g\rightarrow\infty$ is in each case
\beq
\label{exactalpha}
\al=\f{\pi^2}{128}\approx0.0771063.
\eeq
The coefficients $a_l$ and $a_m'$ in the weak- and strong-coupling expansions
(\ref{alw}) and (\ref{als}), respectively, can be obtained to arbitrary
order simply by Taylor-expanding (\ref{ec}).
Note how this implies $q=1$ in VPT.
This is the reason why we used $q=1$ in VPT for the membrane problem.

While for general potentials, the ground state energy cannot be computed
exactly, it is possible to compute all Feynman diagrams analytically
(see appendix \ref{qmintegrals}).
Alternatively, it is possible to compute the coefficients $a_L$ to
arbitrary order by generalizing \cite{KlChHa} the Bender-Wu recursion
relation for the anharmonic oscillator \cite{BeWu}.
The generalized relation reads (correcting some typos in \cite{KlChHa})
\beq
4jc_{nj}=2(j+1)(2j+1)c_{n,j+1}
-\sum_{k=1}^n(-1)^k\ep_{2k+2}c_{n-k,j-k-1}
-2\sum_{k=1}^{n-1}c_{k1}c_{n-k,j},~~~~1\leq j\leq 2n,
\eeq
$c_{00}=1$ and in all other cases $c_{nj}=0$.
The $a_L$ are then given by
\beq
\label{a}
a_L=\left(-\f{1}{4}\right)^Lc_{L-1,1}~~~~L\geq2.
\eeq

\begin{figure}[ht]
\begin{center}
\includegraphics[height=5cm,angle=0]{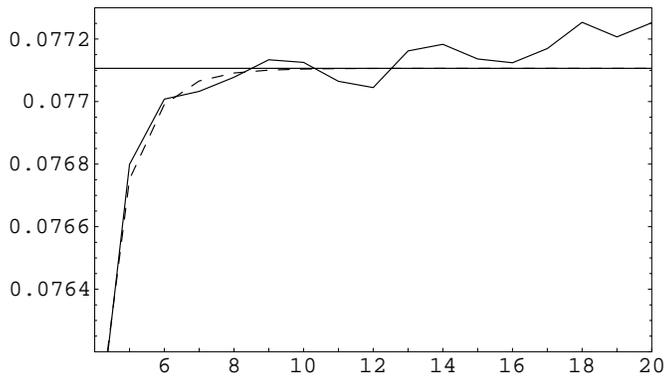}
\end{center}
\caption{\protect\label{qmvarpert1}
Quantum mechanical particle in a box.
$\alpha$ as a function of the loop order $L$ for $q=1$ for $V_c$ (dashed line),
$V_a$ (solid line) and the exact result (horizontal line).
Note how the convergence towards the exact result is exponentially
fast for $V_t$, while questionable for $V_a$.}
\end{figure}

\begin{figure}[ht]
\begin{center}
\begin{tabular}{cc}
\includegraphics[height=5cm,angle=0]{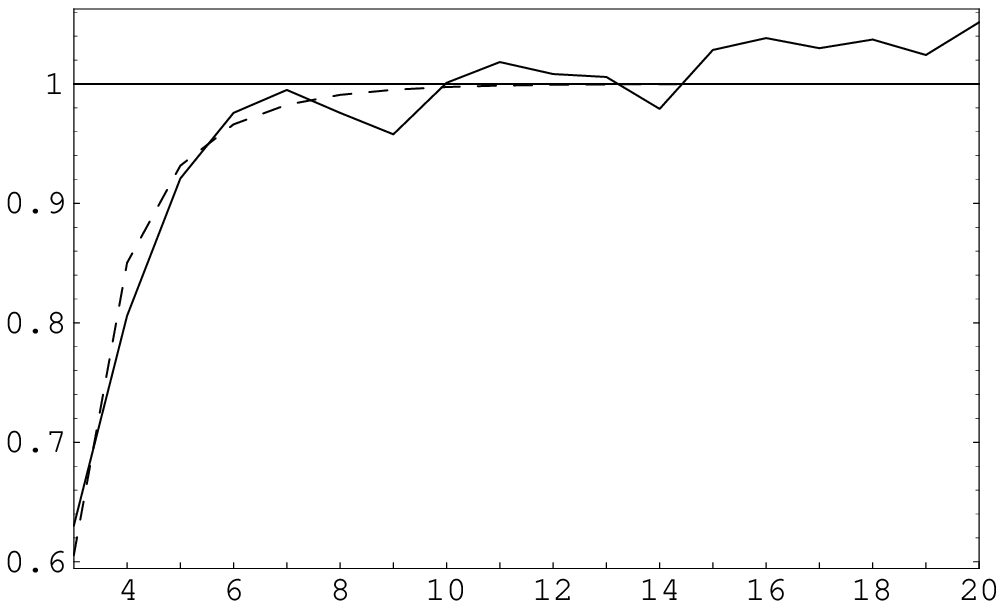}
&~
\includegraphics[height=4.9cm,angle=0]{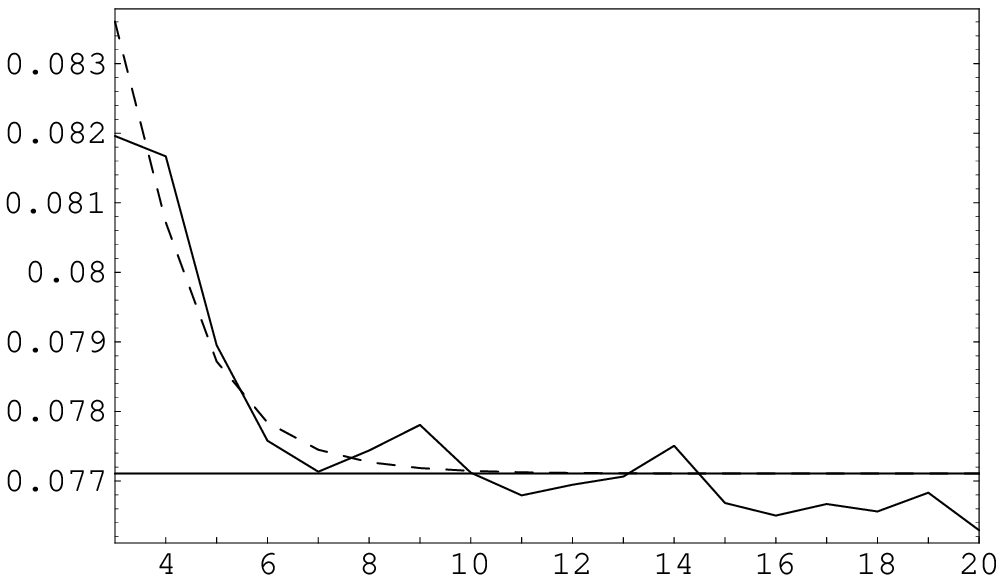}
\\
a & b
\end{tabular}
\end{center}
\caption{\protect\label{qmvarpertq}
Quantum mechanical particle in a box.
(a) Self-consistently determined $q$ as a function of the loop order $L$
from $V_c$ (dashed line), $V_a$ (solid line) and $q=1$.
Note how the convergence towards $q=1$ is exponentially
fast for $V_c$, while for $V_a$, no convergence is obvious, although $q$
is \mbox{around 1}.
(b) $\alpha$ as a function of the loop order with self-consistently
determined $q$ for $V_c$ (dashed line), $V_a$ (solid line) and the exact
result (horizontal line).
Note how the convergence towards the exact result is exponentially
fast for $V_c$, while questionable for $V_a$.}
\end{figure}

The results of carrying out VPT through 20 loops for the potentials $V_c$
and $V_a$ are collected in Table \ref{qmvarperttab} and illustrated in
Figs.\ \ref{qmvarpert1} and \ref{qmvarpertq}.
For fixed $q=1$ we get exponentially fast convergence towards the exact
value of $\al$ for $V_c$.
For $V_a$ no convergence is obvious, although the values obtained through
the order considered are not far from the exact $\al$.
Essentially the same is true when determining $q$ self-consistently,
except that the convergence towards the exact value of $\al$ is delayed
as compared to taking $q=1$.
For $V_c$, $q=1$ is approached exponentially fast, while for $V_a$,
$q>1$ seems preferred at higher orders.

\begin{table}[t]
\begin{center}
\begin{tabular}{c||l||l||l|l||l||l||l|l}
& \multicolumn{4}{c||}{$V_c$} & \multicolumn{4}{c}{$V_a$}
\\\hline
&& \multicolumn{1}{c||}{$q=1$}
& \multicolumn{2}{c||}{$q$ self.-cons.} &
& \multicolumn{1}{c||}{$q=1$}
& \multicolumn{2}{c}{$q$ self.-cons.}
\\\hline
$L$
& \multicolumn{1}{c||}{$a_L$} & \multicolumn{1}{c||}{$\al$} 
& \multicolumn{1}{c|}{$q$} & \multicolumn{1}{c||}{$\al$}
& \multicolumn{1}{c||}{$a_L$} & \multicolumn{1}{c||}{$\al$}
& \multicolumn{1}{c|}{$q$} & \multicolumn{1}{c}{$\al$}
\\\hline
 0 & \pp$0.0506606$ &&&& $\pp0.0416667$ &&& \\
 1 & \pp$0.125$ &&&& $\pp0.125$ &&& \\

 2 & \pp$0.154213$ & $0.0385530$ &&
   & \pp$0.15625$ & $0.0390625$ & $0.309401$ & $$ \\
 3 & \pp$0.0951261$ & $0.0719411$ & $0.605551$ & $0.0836038$
   & \pp$0.0911458$ & $0.0717445$ & $0.630222$ & $0.0819600$ \\
 4 & \pp$0$ & $0.0758821$ & $0.850234$ & $0.0807166$
   & \pp$0.00325521$ & $0.0758318$ & $0.805894$ & $0.0816667$ \\
 5 & $-0.0361959$ & $0.0767518$ & $0.931591$ & $0.0787187$
   & $-0.0340667$ & $0.0767990$ & $0.920850$ & $0.0789522$ \\
 6 & \pp$0$ & $0.0769910$ & $0.966170$ & $0.0778393$
   & $-0.012597$ & $0.0770078$ & $0.975808$ & $0.0775787$ \\
 7 & \pp$0.0275454$ & $0.0770659$ & $0.982590$ & $0.0774492$
   & \pp$0.0421369$ & $0.0770326$ & $0.994979$ & $0.0771334$ \\
 8 & \pp$0$ & $0.0770913$ & $0.990852$ & $0.0772701$
   & \pp$0.0400356$ & $0.0770777$ & $0.975795$ & $0.0774386$ \\
 9 & $-0.0262028$ & $0.0771005$ & $0.995143$ & $0.0771857$
   & $-0.164914$ & $0.0771337$ & $0.957841$ & $0.0778059$ \\
10 & \pp$0$ & $0.0771040$ & $0.997410$ & $0.0771451$
   & $-0.207989$ & $0.0771252$ & $1.000970$ & $0.0771197$ \\
11 & \pp$0.0279168$ & $0.0771054$ & $0.998617$ & $0.0771254$
   & \pp$1.56427$ & $0.0770648$ & $1.018330$ & $0.0767930$ \\
12 & \pp$0$ & $0.0771059$ & $0.999262$ & $0.0771157$
   & \pp$2.21468$ & $0.0770447$ & $1.008300$ & $0.0769460$ \\
13 & $-0.0318674$ & $0.0771061$ & $0.999607$ & $0.0771110$
   & $-25.1291$ & $0.0771620$ & $1.005830$ & $0.0770638$ \\
14 & \pp$0$ & $0.0771062$ & $0.999792$ & $0.0771086$
   & $-43.0543$ & $0.0771831$ & $0.979132$ & $0.0775060$ \\
15 & \pp$0.0381093$ & $0.0771063$ & $0.999890$ & $0.0771074$
   & \pp$585.908$ & $0.0771362$ & $1.028460$ & $0.0766838$ \\
16 & \pp$0$ & $0.0771063$ & $0.999942$ & $0.0771069$
   & \pp$1288.21$ & $0.0771241$ & $1.038400$ & $0.0765030$ \\
17 & $-0.0471274$ & $0.0771063$ & $0.999969$ & $0.0771066$
   & $-18478.5$ & $0.0771695$ & $1.029930$ & $0.0766691$ \\
18 & \pp$0$ & $0.0771063$ & $0.999984$ & $0.0771064$
   & $-53154.3$ & $0.0772534$ & $1.037170$ & $0.0765617$ \\
19 & \pp$0.0597739$ & $0.0771063$ & $0.999992$ & $0.0771064$
   & \pp$753376$ & $0.0772069$ & $1.024240$ & $0.0768319$ \\
20 & \pp$0$ & $0.0771063$ & $0.999996$ & $0.0771063$
   & \pp$2833593$ & $0.0772520$ & $1.051640$ & $0.0762912$ \\
\end{tabular}
\caption{\protect\label{qmvarperttab}
Determination of $\al$ (exact value $\al=0.0771063\ldots$) for QM particle
in a box through 20 loops.
Values of $\al$ for both potentials with $q=1$ and values of $q$ and $\al$
for both potentials with self-consistently determined $q$ are shown.}
\end{center}
\end{table}

It is likely that the inferior convergence behavior for the potential
$V_a$ originates in our missing understanding of the analytical
structure of $E^{(0)}$ as a function of $g$.
It is possible that the strong-coupling behavior is not of the form
(\ref{strongexpansion}) or that the strong-coupling expansion has a zero
radius of convergence.
Numerically, the deviations of the coefficients $a_L$ from those
for $V_c$ are relatively small in low orders, in particular the deviation
from $a_L=0$ for small even $L>2$.
Note how this is very similar to the results of the membrane loop
expansion (\ref{memexpansion}).
Another likely similarity between the membrane problem and the QM problem
with potential $V_a$ is the factorial growth of $a_L$ with $L$ for
large $L$.
In this respect, using $V_c$ in the QM problem is very special, as already
noted in \cite{KlChHa}, and it appears likely that the $a_L$ grow
factorially for the membrane problem for both $V_c$ and $V_a$.
Note, however, that the results for $\al$ with $q=1$ and also for $\al$
with self-consistently determined $q$ improve with increasing $L$ as long
as the $a_L$ do not significantly grow.
We will come back to this point in sec.\ \ref{conclusions}.

\section{Membrane Problem With \bm$\al(g)$ From Particle in a Box}
\label{memqm}
Let us compare the values for the QM expansion coefficients and the
corresponding membrane coefficients using the potential $V_c$:
\bea
\begin{tabular}{c|l|l|l}
$L$&\multicolumn{1}{c|}{$a^{\rm QM}_L$}&
\multicolumn{1}{c|}{$a^{\rm memb}_L$}&
\multicolumn{1}{c}{$a^{\rm memb}_L-a^{\rm QM}_L$}\\\hline
$1$ & $\pp 0.125$     & $\pp 0.125$     & $\pp 0$ \\
$2$ & $\pp 0.154213$  & $\pp 0.154213$  & $\pp 0$ \\
$3$ & $\pp 0.095126$  & $\pp 0.105998$  & $\pp 0.010872$ \\
$4$ & $\pp 0$         & $\pp 0.026569$  & $\pp 0.026569$\\
$5$ & $   -0.036196$  & $   -0.034229$  & $\pp 0.001967$\\
\end{tabular}
\eea
We see that the relative difference through the order considered is small
when both $a_L^{\rm memb}$ and $a_L^{\rm QM}$ are nonzero.
This is the motivation to carry out a different procedure for finding
$\al$ from the loop expansion.
Instead of asking directly what $\alpha$ is for a given potential for the
membrane case, we slightly modify the $\ep_k$ order by order such that the
expansion of $\al(g)$ is identical to that of the QM case with potential
$V_c$ and ask where the resulting potential has the nearest
singularity.
The scaling relation $f\propto1/d^2$ when $m^2=0$ allows us then
to recover $\al$ for the membrane case.

The expansion coefficients of the potentials are
\beq
\begin{tabular}{l|l|l}
& \multicolumn{1}{c|}{QM: $V_c$} & \multicolumn{1}{c}{memb.}
\\\hline
$\ep_0$ & $0.0506606$ & $0.0506606$ \\
$\ep_2$ & $0.5$ & $0.5$ \\
$\ep_4$ & $3.28987$ & $3.28987$ \\
$\ep_6$ & $18.3995$ & $18.0284$ \\
$\ep_8$ & $94.6129$ & $89.5702$ \\
$\ep_{10}$ & $462.545$ & $419.568$ \\
\end{tabular}
\eeq

Let us instead investigate the expansion of
\beq
\label{sqv}
1/\sqrt{2\pi^2V(x)}=\sum_{k=0}^\infty v_{2k}x^{2k},
\eeq
since for the QM case, we have
$1/\sqrt{2\pi^2V(x)}=\cos(\pi x)$.
We can expect a good approximation for the location of the singularity of
$V$ if this singularity is of the quadratic type as in $V_c$ and $V_a$.
The expansion coefficients of the quantity (\ref{sqv}) for the QM and
membrane cases are
\beq
\begin{tabular}{l|l|l}
& \multicolumn{1}{c|}{QM}  & \multicolumn{1}{c}{memb.} \\ \hline
$v_0$    & $\pp 1$         & $\pp 1$       \\
$v_2$    & $   -4.93480$   & $   -4.93480$ \\
$v_4$    & $\pp 4.05871$   & $\pp 4.05871$ \\
$v_6$    & $   -1.33526$   & $\pp 2.32719$ \\
$v_8$    & $\pp 0.235331$  & $   -4.21557$ \\
$v_{10}$ & $   -0.025807$ & $   -0.50636$
\end{tabular}
\eeq
The corresponding zeros $x_0$ of this function, corresponding to the
singularity of $V(x)$, are
\beq
\begin{tabular}{l|l|l}
L & \multicolumn{1}{c|}{QM} & \multicolumn{1}{c}{memb.} \\ \hline
1 & 0.450158 & 0.450158 \\
2 & 0.506893 & 0.506893 \\
3 & 0.499717 & 0.523646 \\
4 & 0.500008 & 0.514714 \\
5 & 0.500000 & 0.514469
\end{tabular}
\eeq
The value of $\al$ in each case is given by $(2x_0)^2$ times the exact
value (\ref{exactalpha}) of $\al$ for QM,
\beq
\begin{tabular}{l|l|l}
L & \multicolumn{1}{c|}{QM} & \multicolumn{1}{c}{memb.} \\ \hline
1 & 0.0625    & 0.0625    \\
2 & 0.0792468 & 0.0792468 \\
3 & 0.0770189 & 0.0845718 \\
4 & 0.0771087 & 0.0817113 \\
5 & 0.0771062 & 0.0816335
\end{tabular}
\eeq
While the correct QM value (\ref{exactalpha}) is approached very
quickly, the convergence in the membrane case is slower.
The fact that the last two values have such a small difference appears
to be accidental.
However, the results point towards a value above $0.080$.

\section{Summary and Discussion}
\label{conclusions}
We have used three methods to extract the pressure exerted by a
tensionless membrane on two infinitely extended parallel walls from a
five-loop calculation for smooth potentials.
While variational perturbation theory with self-consistently determined
$q$ is converging too slowly for quantitative statements at five loops,
variational perturbation theory with the assumption $q=1$ gives a result
$\al\approx0.0813$.
The successive $\al$ values at the various loop orders in
(\ref{alvp}) suggest an extrapolated value of $\al$ between $0.0820$
and $0.0825$.
Fixing $\al(g)$ to resemble the $g$-structure of the ground state energy
of a solvable quantum mechanics problem and analyzing the location of the
singularities next to the origin of the resulting potential leads to
$\al\approx0.0816$.
As opposed to variational perturbation theory with $q=1$, the sequence
given by the considered loop orders gives only a modest indication of
where $\al$ might settle.
The results of both analyses point towards a value above the
Monte Carlo result $\al_{\rm MC}=0.0798\pm0.0003$ \cite{GoKr}.

We have also studied the quantum mechanics problem with a potential
for which we do not know the exact solution but which is very close to
the potential of the solvable problem in the region of interest.
We have investigated both variational perturbation theory with
self-consistently determined $q$ and with $q=1$.
The result is that the exponentially fast convergence of the solvable
model towards the exact result for $\al$ cannot be expected for the
general case, although good estimates of the exact result are obtained.
Since numerically, this case is close to what happens in the membrane case,
this gives us an indication on how trustworthy our results are.
As already noted at the end of sec.\ \ref{QM}, we can roughly state that
the results for $\al$ in the quantum mechanics problem do not improve after
the $a_L$ start to significantly increase in magnitude.
Along this reasoning, one may still expect improving the results for the
membrane problem by proceeding to higher loop orders.
In particular, going to six loops appears feasible with reasonable effort,
since only 8 of the 83 diagrams to be evaluated have no cutvertex and
have therefore a true six-loop topology, as noted in Table \ref{numbers}.

\section*{Acknowledgments}
The author is grateful to H.~Kleinert for many useful discussions and
for suggesting various resummation methods to extract $\al$ from the
loop expansion.

\appendix

\section{Recursion Relation for the Loop Expansion}
\label{loopexp}
Here we define a loop expansion of the free energy and derive recursion
relations for obtaining the required diagrams in a systematic way along
the lines of \cite{recrel}.
We continue to work with $\ka=k_BT=1$.

We write the energy functional (\ref{epenergy}) as
\beq
E[\vp,G,\{L^{(2k)}\}]=\f{1}{2}G^{-1}_{12}\vp_1\vp_2
+L^{(0)}+\sum_{k=2}^\infty L^{(2k)}_{1,\ldots,2k}\vp_1\cdots\vp_{2k}
\eeq
with totally symmetric tensors $G^{-1}$ and $L^{(2k)}$.
Their indices $1,2,\ldots,2k$ are shorthands for space arguments
$\vx_1,\ldots,\vx_{2k}$ and a generalized Einstein convention implies
integration over space arguments that appear twice in a term.
Comparison with (\ref{epenergy}) shows that
\beq
G^{-1}_{12}=\de_{12}[\p_1^2\p_2^2+m^4]
\eeq
and
\beq
L^{(2k)}_{1,\ldots,2k}=m^4\ep_{2k}d^{2(1-k)}\de_{1,\ldots,2k}
\eeq
with
\beq
\de_{1,\ldots,2k}\equiv
\int d^2x\de(\vx-\vx_1)\cdots\de(\vx-\vx_{2k}).
\eeq
Note that the index of $\ep_{2k}$ does not indicate a space argument and
is exempted from the summation convention.

The free energy $Af_m=-W$ is given by
\beq
\exp(W[G,\{L^{(2k)}\}])=\int D\vp\exp(-E[\vp,G,\{L^{(2k)}\}]).
\eeq
$W$ obeys the functional differential equation
\bea
\label{wid}
0
&=&
\int D\vp\f{\de}{\de\vp_1}\{\vp_0\exp(-E[\vp,G,\{L^{(2k)}\}])\}
\nn\\
&=&
\left(\de_{01}
+2G^{-1}_{12}\f{\de}{\de G^{-1}_{02}}
-16L^{(4)}_{1234}\f{\de^2}{\de G^{-1}_{02}\de G^{-1}_{34}}
-4\sum_{k=3}^\infty kL^{(2k)}_{1,\ldots,2k}
\f{\de^2}{\de G^{-1}_{02}\de L^{(2k-2)}_{3,\ldots,2k}}
\right)
\exp(W[G,\{L^{(2k)}\}]).\nn\\
\eea
Splitting $W\equiv W|_{L^{(2k)}=0}+W_I\equiv W_0+W_I$,
so that $W_0$ obeys
\beq
\left(\de_{01}+2G^{-1}_{12}\f{\de}{\de G^{-1}_{02}}
\right)\exp(W_0[G])=0,
\eeq
we get from (\ref{wid})
\beq
\label{wim}
2G^{-1}_{12}\f{\de W_I}{\de G^{-1}_{12}}
-16L^{(4)}_{1234}\left(
\f{\de^2W}{\de G^{-1}_{12}\de G^{-1}_{34}}
+\f{\de W}{\de G^{-1}_{12}}\f{\de W}{\de G^{-1}_{34}}\right)
-4\sum_{k=3}^\infty kL^{(2k)}_{1,\ldots,2k}
\left(\f{\de^2W}{\de G^{-1}_{12}\de L^{(2k-2)}_{3,\ldots,2k}}
+\f{\de W}{\de G^{-1}_{12}}
\f{\de W}{\de L^{(2k-2)}_{3,\ldots,2k}}\right)=0,
\eeq
where we additionally have identified indices $0$ and $1$ and
integrated over the respective variable.
With
\beq
\f{\de W_0}{\de G^{-1}_{12}}=-\f{1}{2}G_{12}
\eeq
and
\beq
\f{\de^2 W_0}{\de G^{-1}_{12}\de G^{-1}_{34}}
=\f{1}{4}(G_{13}G_{24}+G_{14}G_{23}),
\eeq
Eq.\ (\ref{wim}) may be transformed into
\bea
\label{wieq}
G_{12}\f{\de W_I}{\de G_{12}}
&=&
-6L^{(4)}_{1234}G_{12}G_{34}
-24L^{(4)}_{1234}G_{12}G_{35}G_{46}\f{\de W_I}{\de G_{56}}
-8L^{(4)}_{1234}G_{15}G_{26}G_{37}G_{48}\left(
\f{\de^2W_I}{\de G_{56}\de G_{78}}
+\f{\de W_I}{\de G_{56}}\f{\de W_I}{\de G_{78}}\right)
\nn\\
&&{}
+\sum_{k=3}^\infty kL^{(2k)}_{1,\ldots,2k}
\left[G_{12}\f{\de W_I}{\de L^{(2k-2)}_{3,\ldots,2k}}
+2G_{1\bar{1}}G_{2\bar{2}}
\left(\f{\de^2W_I}{\de G_{\bar{1}\bar{2}}\de L^{(2k-2)}_{3,\ldots,2k}}
+\f{\de W_I}{\de G_{\bar{1}\bar{2}}}
\f{\de W_I}{\de L^{(2k-2)}_{3,\ldots,2k}}\right)\right].
\eea
Define a loop expansion
\beq
W=\sum_{L=0}^\infty W^{(L)}
\eeq
and set
\beq
W^{(0)}=-\f{A\ep_0}{d^2g^2}
\eeq
by an appropriate normalization
of the path integral measure $D\vp$.
Then
\beq
W_0=W^{(1)}=-\f{1}{2}\ln(G^{-1})_{11}
\eeq
and
\beq
~~~~W_I=\sum_{L=2}^\infty W^{(L)}.
\eeq
Eq.\ (\ref{wieq}) separates into the two-loop equation
\beq
\label{w2eq}
G_{12}\f{\de W^{(2)}}{\de G_{12}}
+6L^{(4)}_{1234}G_{12}G_{34}=0
\eeq
and the recursion relation
\bea
\label{wirr}
G_{12}\f{\de W^{(L)}}{\de G_{12}}
&=&
-24L^{(4)}_{1234}G_{12}G_{35}G_{46}
\f{\de W^{(L-1)}}{\de G_{56}}
-8L^{(4)}_{1234}G_{15}G_{26}G_{37}G_{48}\left(
\f{\de^2W^{(L-1)}}{\de G_{56}\de G_{78}}
+\sum_{l=2}^{L-2}
\f{\de W^{(l)}}{\de G_{56}}\f{\de W^{(L-l)}}{\de G_{78}}\right)
\nn\\
&&{}
+\sum_{k=3}^LkL^{(2k)}_{1,\ldots,2k}
\left[G_{12}\f{\de W^{(L-1)}}{\de L^{(2k-2)}_{3,\ldots,2k}}
+2G_{1\bar{1}}G_{2\bar{2}}
\left(\f{\de^2W^{(L-1)}}{\de G_{\bar{1}\bar{2}}\de L^{(2k-2)}_{3,\ldots,2k}}
+\sum_{l=2}^{L-k+1}\f{\de W^{(l)}}{\de G_{\bar{1}\bar{2}}}
\f{\de W^{(L-l)}}{\de L^{(2k-2)}_{3,\ldots,2k}}\right)\right],
\eea
which holds for $L>2$ and where we have taken into account that a
diagram containing the tensor $L^{(2k)}$ has at least $k$ loops.

Eq.\ (\ref{w2eq}) is solved by
\beq
\label{w2}
W^{(2)}=-3L^{(4)}_{1234}G_{12}G_{34}.
\eeq
Before carrying out the recursion relation (\ref{wirr}), let us
introduce a graphical representation of the resulting terms.
Represent each free propagator $G_{12}$ by a line with two ends
corresponding to the two space arguments $\vx_1$ and $\vx_2$
and each tensor $-L^{(n)}_{1,\ldots,n}$ by a dot.
Each line end is connected to the dot with an identical space
argument.
Then a dot corresponding to a tensor $L^{(n)}_{1,\ldots,n}$
has $n$ line ends connected to it.
In this way all terms appearing in the $W^{(L)}$ with $L\neq1$ can
be graphically represented.
The zero-loop order is represented by a dot without lines,
\beq
W^{(0)}=
\rule[-2pt]{0pt}{10pt}
\bpi(10,0)
\put(5,3){\circle*{4}}
\epi.
\eeq
Only $W^{(1)}$ does not fit into the graphical scheme above and as usual
we use the graphical representation
\beq
W^{(1)}=\f{1}{2}
\rule[-10pt]{0pt}{26pt}
\bpi(26,0)
\put(13,3){\circle{16}}
\epi
\eeq
for it.
Now we may write (\ref{w2}) as
\beq
W^{(2)}=3
\rule[-10pt]{0pt}{26pt}
\bpi(42,0)
\put(13,3){\circle{16}}
\put(29,3){\circle{16}}
\put(21,3){\circle*{4}}
\epi,
\eeq
which is the starting point for the recursive determination of the
other $W^{(L)}$.
For instance, the three-loop contribution to $W$ is
\beq
W^{(3)}=12
\rule[-14pt]{0pt}{34pt}
\bpi(34,0)
\put(17,3){\circle{24}}
\put(17,3){\oval(24,8)}
\put(5,3){\circle*{4}}
\put(29,3){\circle*{4}}
\epi
+36
\rule[-10pt]{0pt}{26pt}
\bpi(58,0)
\put(13,3){\circle{16}}
\put(29,3){\circle{16}}
\put(45,3){\circle{16}}
\put(21,3){\circle*{4}}
\put(37,3){\circle*{4}}
\epi
+15
\rule[-13pt]{0pt}{37pt}
\bpi(42,0)
\put(21,3){\circle*{4}}
\qbezier(21,3)(13.79,10.21)(13.79,13.79)
\qbezier(13.79,13.79)(13.79,21.)(21.,21.)
\qbezier(21,3)(28.21,10.21)(28.21,13.79)
\qbezier(28.21,13.79)(28.21,21.)(21.,21.)
\qbezier(21,3)(18.36,-6.85)(15.26,-8.64)
\qbezier(15.26,-8.64)(9.02,-12.24)(5.41,-6.)
\qbezier(21,3)(11.15,5.64)(8.05,3.85)
\qbezier(8.05,3.85)(1.81,0.24)(5.41,-6.)
\qbezier(21,3)(30.85,5.64)(33.95,3.85)
\qbezier(33.95,3.85)(40.19,0.24)(36.59,-6.)
\qbezier(21,3)(23.64,-6.85)(26.74,-8.64)
\qbezier(26.74,-8.64)(32.98,-12.24)(36.59,-6.)
\epi.
\eeq
In Table \ref{numbers} we list the numbers of different diagrams through
seven loops and the numbers of diagrams at each loop order which have no
cutvertex (by definition, upon cutting through such a vertex appropriately,
a diagram decomposes into two diagrams, which consequently have independent
momentum integrations) and have therefore their full loop topology.
The contributions through five loops for both the QM and the membrane
problem are collected in Table \ref{diagrams}.

\begin{table}[t]
\begin{center}
\begin{tabular}{c||c|c|c|c|c|c|c}
number of loops $L$ & 1 & 2 & 3 & 4 & 5 & 6 & 7\\\hline\hline
diagrams & 1 & 1 & 3 & 7 & 24 & 83 & 376\\\hline
diagrams with $L$-loop topology & 1 & 0 & 1 & 1 & 5 & 8 & 37
\end{tabular}
\end{center}
\caption{\label{numbers}
Numbers of vacuum diagrams for some low loop orders and numbers of those
with full loop topology.}
\label{numbersofdiagrams}
\end{table}

\section{Evaluation of Quantum Mechanic Integrals}
\label{qmintegrals}
All integrals in the QM case can be evaluated analytically.
The propagator reads
\beq
\De(k^2,m^2)=\f{1}{k^2+m^2},
\eeq
its Fourier transform is
\beq
\tilde{\De}(t,m)=\int_{-\infty}^{+\infty}\f{dk}{2\pi}\f{e^{ikt}}{k^2+m^2}
=\f{e^{-m|t|}}{2m}.
\eeq
For most integrals it is convenient to work in $t$-space, omitting
the last $t$-integration (which, due to time translation invariance, gives
a factor $T$, the total interaction time).
An exception are the one-loop integrals, which are computed easiest in
momentum space.
Using dimensional regularization, we get
\beq
J_0'
=\left.\int\f{d^Dp}{(2\pi)^D}\ln(p^2+m^2)\right|_{D=1}
=\left.-\f{1}{D}\int\f{d^Dp}{(2\pi)^D}p_\mu\f{\p}{\p_\mu}\ln(p^2+m^2)
\right|_{D=1}
=\left.-\f{2}{D}\int\f{d^Dp}{(2\pi)^D}\f{p^2}{p^2+m^2}\right|_{D=1}
=m
\eeq
for the only diverging integral, while the other one-loop integrals are
given by
\beq
J_n\equiv\int_{-\infty}^{+\infty}\f{dp}{2\pi}\f{1}{(p^2+m^2)^n}
=\f{m^{1-2n}}{2}\f{\Ga(n-\f{1}{2})}{\sqrt{\pi}\Ga(n)}.
\eeq
Note the similarity to the membrane one-loop integrals (\ref{j0}) and
(\ref{oneloop}) below, so that
\beq
\label{qmmbsim}
{J_0'}^{\rm qm}=4m^{-1}{J_0'}^{\rm memb},~~~~~~
J_n^{\rm qm}=4m^{2n-1}J_n^{\rm memb}.
\eeq

\section{Evaluation of Membrane Integrals}
\label{mbintegrals}
In the following, we will always assume that $m^2>0$.
We give results in a form suited for numerical integration of the
remaining loop momenta.

\subsection{One-Loop Vacuum Integrals}
The ubiquitous one-loop integrals without external momenta can be computed
analytically as
\beq
\label{j0}
J_0'\equiv
\left.\int\f{d^Dp}{(2\pi)^D}\ln(p^4+m^4)\right|_{D=2}
=\left.-\f{1}{D}\int\f{d^Dp}{(2\pi)^D}p_\mu\f{\p}{\p_\mu}\ln(p^4+m^4)
\right|_{D=2}
=\left.-\f{4}{D}\int\f{d^Dp}{(2\pi)^D}\f{p^4}{p^4+m^4}\right|_{D=2}
=\f{m^2}{4}
\eeq
and
\beq
\label{oneloop}
J_n\equiv\int_p\f{1}{(p^4+m^4)^n}
=\f{m^{2-4n}}{8}\f{\Ga(n-\f{1}{2})}{\sqrt{\pi}\Ga(n)},
\eeq
where dimensional regularization has been employed for $J_0'$.

\subsection{One-Loop Bubble}
\label{oneloopsct}
Several diagrams contain the one-loop bubble
\bea
\label{De1}
\lefteqn{\De_{\rm ol}^{(1,1)}(k^2,m^2)=
\rule[-10pt]{0pt}{26pt}
\bpi(34,0)
\put(5,3){\oval(8,8)[r]}
\put(9,3){\circle*{4}}
\put(17,3){\circle{16}}
\put(25,3){\circle*{4}}
\put(29,3){\oval(8,8)[l]}
\epi
=\int_p\De((k+p)^2,m^2)\De(p^2,m^2)}
\nn\\
&=&
-\f{1}{4m^4}\int_p
\left(\f{1}{(k+p)^2+im^2}-\f{1}{(k+p)^2-im^2}\right)
\left(\f{1}{p^2+im^2}-\f{1}{p^2-im^2}\right)
\nn\\
&=&
-\f{1}{4m^4}\left[\psi_{\rm ol}^{(1,1)}(k^2,m^2,m^2)
-\psi_{\rm ol}^{(1,1)}(k^2,m^2,-m^2)-\psi_{\rm ol}^{(1,1)}(k^2,-m^2,m^2)
+\psi_{\rm ol}^{(1,1)}(k^2,-m^2,-m^2)\right]
\nn\\
\eea
with $\psi_{\rm ol}^{(1,1)}$ defined by
\bea
\psi_{\rm ol}^{(1,1)}(k^2,m_1^2,m_2^2)
&\equiv&
\int_p\f{1}{[(p+k)^2+im_1^2](p^2+im_2^2)}.
\eea
Elementary integration gives
\beq
\psi_{\rm ol}^{(1,1)}(k^2,\pm m^2,\pm m^2)
=\f{1}{2\pi\sqrt{k^2}\sqrt{k^2\pm4im^2}}
\ln\f{\sqrt{k^2\pm4im^2}+\sqrt{k^2}}{\sqrt{k^2\pm4im^2}-\sqrt{k^2}}
\eeq
and
\beq
\psi_{\rm ol}^{(1,1)}(k^2,\pm m^2,\mp m^2)
=\f{1}{4\pi\sqrt{k^4-4m^4}}
\ln\f{k^2+\sqrt{k^4-4m^4}}{k^2-\sqrt{k^4-4m^4}},
\eeq
so that
\beq
\label{deol11}
\De_{\rm ol}^{(1,1)}(k^2,m^2)
=\f{1}{8\pi m^4}\left[\f{1}{\sqrt{k^4-4m^4}}
\ln\f{k^2+\sqrt{k^4-4m^4}}{k^2-\sqrt{k^4-4m^4}}
-2{\rm Re}\left(\f{1}{\sqrt{k^2}\sqrt{k^2-4im^2}}
\ln\f{\sqrt{k^2-4im^2}+\sqrt{k^2}}{\sqrt{k^2-4im^2}-\sqrt{k^2}}
\right)\right].
\eeq
Now we can also easily compute
\bea
\label{deol12}
\De_{\rm ol}^{(1,2)}(k^2,m^2)
&=&
\rule[-10pt]{0pt}{26pt}
\bpi(34,0)
\put(5,3){\oval(8,8)[r]}
\put(17,3){\circle{16}}
\put(9,3){\circle*{4}}
\put(17,11){\circle*{4}}
\put(25,3){\circle*{4}}
\put(29,3){\oval(8,8)[l]}
\epi
=\int_p\De((k+p)^2,m^2)\De(p^2,m^2)^2
=-\f{1}{4m^2}\f{\p}{\p m^2}\De_{\rm ol}^{(1,1)}(k^2,m^2)
\nn\\
&=&
\f{5k^2}{4\pi m^4(k^4-4m^4)(k^4+16m^4)}
\nn\\
&&{}
+\f{1}{16\pi m^8}
\left[
\f{k^4-6m^4}{(k^4{-}4m^4)^{3/2}}
\ln\f{k^2+\sqrt{k^4{-}4m^4}}{k^2-\sqrt{k^4{-}4m^4}}
-2{\rm Re}\left(
\f{k^2+5im^2}{\sqrt{k^2}(k^2{+}4im^2)^{3/2}}
\ln\f{\sqrt{k^2{+}4im^2}+\sqrt{k^2}}{\sqrt{k^2{+}4im^2}-\sqrt{k^2}}
\right)\right].
\nn\\
\eea

\subsection{Sunset Self-Energy}
\label{sunsetsct}
For several diagrams we need to compute
\bea
\label{ssse}
\lefteqn{\De_{\rm ss}(k^2,m^2)=
\rule[-10pt]{0pt}{26pt}
\bpi(38,0)
\put(5,3){\line(1,0){28}}
\put(19,3){\circle{16}}
\put(11,3){\circle*{4}}
\put(27,3){\circle*{4}}
\epi
=\int_{pq}\De((k+p)^2,m^2)\De((p+q)^2,m^2)\De(q^2,m^2)}
\nn\\
&=&
-\f{i}{8m^6}\left[
\psi_{\rm ss}(k^2,m^2,m^2,m^2)-\psi_{\rm ss}(k^2,m^2,m^2,-m^2)
-\psi_{\rm ss}(k^2,m^2,-m^2,m^2)+\psi_{\rm ss}(k^2,m^2,-m^2,-m^2)
\right.
\nn\\
&&~~~~~~~~~~{}
\left.
-\psi_{\rm ss}(k^2,-m^2,m^2,m^2)+\psi_{\rm ss}(k^2,-m^2,m^2,-m^2)
+\psi_{\rm ss}(k^2,-m^2,-m^2,m^2)-\psi_{\rm ss}(k^2,-m^2,-m^2,-m^2)\right]
\nn\\
\eea
with $\psi_{\rm ss}$ defined by
\beq
\psi_{\rm ss}(k^2,m_1^2,m_2^2,m_3^2)
\equiv\int_{pq}\f{1}{[(k+p)^2+im_1^2][(p+q)^2+im_2^2](q^2+im_3^2)}.
\eeq
Note that
\beq
\label{psiids1}
\psi_{\rm ss}(k^2,m^2,m^2,m^2)=\psi_{\rm ss}(k^2,-m^2,-m^2,-m^2)^*
\eeq
and
\bea
\label{psiids2}
\lefteqn{\psi_{\rm ss}(k^2,-m^2,m^2,m^2)
=\psi_{\rm ss}(k^2,m^2,-m^2,m^2)=\psi_{\rm ss}(k^2,m^2,m^2,-m^2)}
\nn\\
&=&
\psi_{\rm ss}(k^2,m^2,-m^2,-m^2)^*
=\psi_{\rm ss}(k^2,-m^2,m^2,-m^2)^*=\psi_{\rm ss}(k^2,-m^2,-m^2,m^2)^*.
\eea
$\psi_{\rm ss}$ may be evaluated as
\bea
\psi_{\rm ss}(k^2,m_1^2,m_2^2,m_3^2)
&=&
\int_0^1d\al\int_p\f{1}{(k+p)^2+im_1^2}
\int_q\f{1}{[q^2+2\al p\cdot q+\al p^2+\al im_2^2+(1-\al)im_3^2]^2}
\nn\\
&=&
\f{1}{4\pi}\int_0^1d\al\int_p
\f{1}{[(k+p)^2+im_1^2][\al(1-\al)p^2+\al im_2^2+(1-\al)im_3^2]}
\nn\\
&=&
\f{1}{4\pi}\int_0^1\f{d\al}{\al(1-\al)}\int_0^1d\be
\int_p
\f{1}{\left[p^2+2(1-\be)p\cdot k(1-\be)(k^2+im_1^2)
+\be i\left(\f{m_2^2}{1-\al}+\f{m_3^2}{\al}\right)\right]^2}
\nn\\
&=&
\f{1}{(4\pi)^2}\int_0^1\f{d\al}{\al(1-\al)}\int_0^1d\be
\f{1}{\be(1-\be)k^2+(1-\be)im_1^2
+\be i\left(\f{m_2^2}{1-\al}+\f{m_3^2}{\al}\right)}
\nn\\
&=&
\f{1}{(4\pi)^2}\int_0^1d\al\int_0^1d\be
\f{1}{\al(1-\al)(1-\be)(\be k^2+im_1^2)
+\be i[\al m_2^2+(1-\al)m_3^2]}
\nn\\
&=&
\f{2}{(4\pi)^2}\int_0^1d\be\int_{-1}^{+1}dx
\f{1}{(1-x^2)(1-\be)(\be k^2+im_1^2)+2i\be[(1+x)m_2^2+(1-x)m_3^2]}.
\eea
If $m_3^2=m_2^2$, this becomes
\bea
\psi_{\rm ss}(k^2,m_1^2,m_2^2,m_2^2)
&=&
\f{2}{(4\pi)^2}\int_0^1d\be\int_{-1}^{+1}dx
\f{1}{(1-x^2)(1-\be)(\be k^2+im_1^2)+4i\be m_2^2},
\nn\\
&=&
\f{2}{(4\pi)^2}\int_0^1\f{d\be}{(1-\be)(\be k^2+im_1^2)}\int_{-1}^{+1}dx
\f{1}{1+\f{4i\be m_2^2}{(1-\be)(\be k^2+im_1^2)}-x^2},
\eea
while for $m_3^2=-m_2^2$ we have
\bea
\psi_{\rm ss}(k^2,m_1^2,m_2^2,-m_2^2)
&=&
\f{2}{(4\pi)^2}\int_0^1d\be\int_{-1}^{+1}dx
\f{1}{(1-x^2)(1-\be)(\be k^2+im_1^2)+4ix\be m_2^2},
\nn\\
&=&
\f{2}{(4\pi)^2}\int_0^1\f{d\be}{(1-\be)(\be k^2+im_1^2)}\int_{-1}^{+1}dx
\f{1}{1-x^2+\f{4i\be m_2^2}{(1-\be)(\be k^2+im_1^2)}x}.
\eea
Noting that the formulas
\beq
f_1(z)=\int_{-1}^{+1}\f{dx}{z-x^2}
=\f{1}{2\sqrt{z}}\int_{-1}^{+1}dx
\left(\f{1}{\sqrt{z}+x}+\f{1}{\sqrt{z}-x}\right)
=\f{\ln(\sqrt{z}+1)-\ln(\sqrt{z}-1)}{\sqrt{z}}
\eeq
and
\bea
f_2(z)
&\equiv&
\int_{-1}^{+1}\f{dx}{1-x^2+2zx}
=\f{1}{2\sqrt{1+z^2}}\int_{-1}^{+1}dx
\left(\f{1}{\sqrt{1+z^2}-z+x}+\f{1}{\sqrt{1+z^2}+z-x}\right)
\nn\\
&=&
\f{1}{2\sqrt{1+z^2}}
\Big[\ln(\sqrt{1+z^2}-z+1)-\ln(\sqrt{1+z^2}-z-1)
-\ln(\sqrt{1+z^2}+z-1)+\ln(\sqrt{1+z^2}+z+1)\Big]~~~~~~~~
\eea
are numerically safe to use if the branch cut of the logarithm is taken
from $0$ to $-\infty$, we may now express the $\psi_{\rm ss}$ in
(\ref{ssse}) with the help of $\be$-integrals involving $f_1$ and $f_2$.
However, these integrals are difficult to evaluate numerically if
$m_3^2=m_2^2=-m_1^2$.
These cases may be avoided by making use of (\ref{psiids2}).
Together with (\ref{psiids1}) we get
\bea
\label{dess}
\De_{\rm ss}(k^2,m^2)
&=&
\f{1}{8m^6}{\rm Im}
[2\psi_{\rm ss}(k^2,m^2,m^2,m^2)-6\psi_{\rm ss}(k^2,m^2,m^2,-m^2)]
\nn\\
&=&
\f{1}{32\pi^2m^6}{\rm Im}\int_0^1\f{d\be}{(1{-}\be)(\be k^2+im^2)}
\left[f_1\left(1+\f{4i\be m^2}{(1{-}\be)(\be k^2+im^2)}\right)
-3f_2\left(\f{2i\be m^2}{(1{-}\be)(\be k^2+im^2)}\right)\right].
\nn\\
\eea
This form is easy to implement numerically.

\subsection{Eye Bubble and Triangle Coupling}
\label{eyebubblesct}
For the integral $I_{5-3}$, we need the eye bubble subdiagram
\bea
\label{deeye}
\De_{\rm eye}(k^2,m^2)
&=&
\rule[-10pt]{0pt}{26pt}
\bpi(50,0)
\put(5,3){\oval(8,8)[r]}
\put(9,3){\circle*{4}}
\put(25,3){\circle{16}}
\put(41,3){\circle*{4}}
\put(25,11){\circle*{4}}
\put(25,-5){\circle*{4}}
\put(25,3){\oval(32,16)}
\put(45,3){\oval(8,8)[l]}
\epi
=
\int_r\De_{\rm tr}(k,r,m^2)^2,
\eea
where the triangle subdiagram $\De_{\rm tr}$ is given by
\bea
\label{detr}
\De_{\rm tr}(k,r,m^2)
&=&
\rule[-8pt]{0pt}{38pt}
\bpi(69,0)
\put(5,-3){\line(1,1){28}}
\put(49,-3){\line(-1,1){28}}
\put(11,3){\line(1,0){32}}
\put(11,3){\circle*{4}}
\put(43,3){\circle*{4}}
\put(27,19){\circle*{4}}
\put(8,10){$p$}
\put(42,10){$p{+}k$}
\put(18,-7){$p{+}r$}
\put(5,-3){\vector(1,1){16}}
\put(21,25){\vector(1,-1){16}}
\put(43,3){\vector(-1,0){18}}
\epi
=
\int_p\De(p^2,m^2)\De((p+k)^2,m^2)\De((p+r)^2,m^2)
\nn\\
&=&
-\f{i}{8(2\pi)^2m^6}\left[\psi_{\rm tr}(k,r,m^2,m^2,m^2)
-\psi_{\rm tr}(k,r,m^2,m^2,-m^2)\right.
\nn\\
&&~~~~~~~~~~~~~~~~{}
-\psi_{\rm tr}(k,r,m^2,-m^2,m^2)
+\psi_{\rm tr}(k,r,m^2,-m^2,-m^2)
\nn\\
&&~~~~~~~~~~~~~~~~{}
-\psi_{\rm tr}(k,r,-m^2,m^2,m^2)
+\psi_{\rm tr}(k,r,-m^2,m^2,-m^2)
\nn\\
&&~~~~~~~~~~~~~~~\left.
{}+\psi_{\rm tr}(k,r,-m^2,-m^2,m^2)
-\psi_{\rm tr}(k,r,-m^2,-m^2,-m^2)\right]
\nn\\
&=&
\f{1}{(4\pi)^2m^6}{\rm Im}\left[\psi_{\rm tr}(k,r,m^2,m^2,m^2)
-\psi_{\rm tr}(k,r,m^2,m^2,-m^2)\right.
\nn\\
&&~~~~~~~~~~~~~~~\left.{}
-\psi_{\rm tr}(k,r,m^2,-m^2,m^2)
+\psi_{\rm tr}(k,r,m^2,-m^2,-m^2)\right]
\eea
with $\psi_{\rm tr}$ defined by
\bea
\label{psitr}
\lefteqn{\psi_{\rm tr}(k,r,m_1^2,m_2^2,m_3^2)\equiv
\int\f{d^2p}{(p^2+im_1^2)[(p+k)^2+im_2^2][(p+r)^2+im_3^2]}}
\nn\\
&=&
\int_0^\infty dpp\int_0^{2\pi}d\phi
\f{1}{(p^2+im_1^2)[p^2+k^2+2pk\cos(\phi-\phi_k)+im_2^2]
[p^2+r^2+2pr\cos(\phi-\phi_r)+im_3^2]}
\nn\\
&=&
\int_0^{2\pi}d\phi\int_0^\infty
\f{dpp}{(p-a_+)(p-a_-)(p-b_+)(p-b_-)(p-c_+)(p-c_-)}
\nn\\
&=&
-\int_0^{2\pi}\!\!d\phi
\Bigg\{\f{1}{a_+{-}a_-}\left[
\f{a_+\ln a_+}{(a_+{-}b_+)(a_+{-}b_-)(a_+{-}c_+)(a_+{-}c_-)}
-\f{a_-\ln a_-}{(a_-{-}b_+)(a_-{-}b_-)(a_-{-}c_+)(a_-{-}c_-)}\right]
\nn\\
&&~~~~~~~~~~~~{}
+\f{1}{b_+{-}b_-}
\left[\f{b_+\ln b_+}{(b_+{-}a_+)(b_+{-}a_-)(b_+{-}c_+)(b_+{-}c_-)}
-\f{b_-\ln b_-}{(b_-{-}a_+)(b_-{-}a_-)(b_-{-}c_+)(b_-{-}c_-)}\right]
\nn\\
&&~~~~~~~~~~~~{}
+\f{1}{c_+{-}c_-}
\left[\f{c_+\ln c_+}{(c_+{-}a_+)(c_+{-}a_-)(c_+{-}b_+)(c_+{-}b_-)}
-\f{c_-\ln c_-}{(c_-{-}a_+)(c_-{-}a_-)(c_-{-}b_+)(c_-{-}b_-)}\right]
\Bigg\}
\eea
with
\beq
a_\pm=\pm i\sqrt{im_1^2},
\eeq
\beq
b_\pm=-k\cos(\phi-\phi_k)\pm i\sqrt{k^2\sin^2(\phi-\phi_k)+im_2^2},
\eeq
\beq
c_\pm=-r\cos(\phi-\phi_r)\pm i\sqrt{r^2\sin^2(\phi-\phi_r)+im_3^2}.
\eeq

For the numerical evaluation of $I_{5-3}$, it is also useful to know the
large-$k$ behavior
\beq
\label{deeyeasymp}
\De_{\rm eye}(k^2,m^2)=\f{2}{k^8}I_{3-1},~~~~~~k^2\gg m^2
\eeq
with $I_{3-1}$ from Table \ref{diagrams}.

Let us remark here that $\psi_{\rm tr}$ and therefore $\De_{\rm tr}$ can
also be computed analytically (e.g.\ by performing the integrals over the
cartesian components $p_x$ and $p_y$ of $p$), but the resulting expressions
are rather lengthy and it may be a delicate issue to remain on the same
Riemann sheet while evaluating the logarithms and square roots involved.

\subsection{Numerical Considerations}
Most integrals are evaluated rather easily.
They are either known analytically, involve only one numeric integration
or involve one such integration involving the function
$\De_{\rm ss}(k^2,m^2)$, whose evaluation implies one numeric integration,
see (\ref{dess}).
These cases are easily dealt with by using any standard software
integration package, e.g.\ {\sc mathematica}, which was used here.
A lot of integrals can be evaluated in different ways, so the safety
of using $\De_{\rm ss}(k^2,m^2)$ inside an integral can be and has
been checked.
For instance, the integral $I_{3-1}$ can be performed by integrating either
$\De_{\rm ol}^{(1,1)}(k^2,m^2)^2$ or $\De_{\rm ss}(k^2,m^2)\De(k^2,m^2)$
over $k^2$.
By such cross checks and by varying the settings within {\sc mathematica}
for the numerical integrations, one may easily gain confidence that the
integrations performed are accurate through the number of digits given in
Table \ref{diagrams}.

The only exception to the above considerations through five loops is
integral $I_{5-3}$.
As indicated in Table \ref{diagrams}, the product 
$\De_{\rm ol}^{(1,1)}(k^2,m^2)\De_{\rm eye}(k^2,m^2)$ has to be
integrated over $k^2$.
While  $\De_{\rm ol}^{(1,1)}(k^2,m^2)$ is known analytically from
(\ref{deol11}), $\De_{\rm eye}(k^2,m^2)$ involves a further
two-dimensional numeric integration of $\De_{\rm tr}(k,r,m^2)^2$ over
$r$, see (\ref{deeye}).
$\De_{\rm tr}(k,r,m^2)$ itself implies a one-dimensional numeric
integration, see (\ref{detr}) and (\ref{psitr}).
Using this method, an uncertainty in the last digit given in Table
\ref{diagrams} remained in our computations.
We achieved the precision through the last digit given in Table
\ref{diagrams} by rewriting $I_{5-3}$ as the five-dimensional integral
\bea
I_{5-3}
&=&
\f{1}{2^7\pi^5}\int_0^\infty dp^2\int_0^\infty dq^2\int_0^\infty dr^2
\int_0^\pi d\phi_{pr}\int_0^{2\pi}d\phi_{qr}
\nn\\
&&{}\times
\De(p^2,m^2)\De(q^2,m^2)\De((p+r)^2,m^2)\De((q+r)^2,m^2)
\De_{\rm ol}^{(1,1)}(r^2,m^2)\De_{\rm ol}^{(1,1)}((p-q)^2,m^2)
\eea
in an obvious notation and asking {\sc mathematica} for successively
increasing precisions of the final result.

The precision achieved for all integrals is several orders of magnitude
better than needed to determine, say, the first three non-zero digits of our
estimates for $\al$.
Higher precision is easily attained for all integrals but $I_{5-3}$, but is
unnecessary for the purpose of this work.

\newpage
\input{mem5paptable}

\end{document}

%% file: mem5paptable.tex
\begin{longtable}[c]{|c|c|c|c|c|rl|}
\caption{Diagrams $L{-}n$ ($n$th $L$-loop diagram) through five loops and
their combinatorial factors $c_{L-n}$, coupling constant factors $g_{L-n}$
and values $I_{L-n}$ of the corresponding integrals for $m=1$.
$D=1$ corresponds to the QM problem and $D=2$ to the membrane problem.
\label{diagrams}}
\\\hline
\rule[-4pt]{0pt}{14pt}$L{-}n$
&diagram&$c_{L-n}$&$g_{L-n}$&$I_{L-n}$ for $D=1$&
\multicolumn{2}{c|}{$I_{L-n}$ for $D=2$}
\\\hline\hline
\endfirsthead
\caption{(continued)}
\\\hline
\rule[-4pt]{0pt}{14pt}$L{-}n$
&diagram&$c_{L-n}$&$g_{L-n}$&$I_{L-n}$ for $D=1$&
\multicolumn{2}{c|}{$I_{L-n}$ for $D=2$}
\\\hline\hline
\endhead
0-1 &
\rule[-4pt]{0pt}{14pt}
\bpi(10,0)
\put(5,3){\circle*{4}}
\epi
& $1$ & $-\ep_0$ & $1$ & $1$ &
\\\hline\hline
1-1 &
\rule[-10pt]{0pt}{26pt}
\bpi(26,0)
\put(13,3){\circle{16}}
\epi
& $1/2$ & $1$ & $-J_0'=-1$ & $-J_0'=$ & $-1/4$
\\\hline\hline
2-1 &
\rule[-10pt]{0pt}{26pt}
\bpi(42,0)
\put(13,3){\circle{16}}
\put(29,3){\circle{16}}
\put(21,3){\circle*{4}}
\epi
& $3$ & $-\ep_4$ & $1/4$
& $J_1^2=$
& $1/64$
\\\hline\hline
3-1 &
\rule[-14pt]{0pt}{34pt}
\bpi(34,0)
\put(17,3){\circle{24}}
\put(17,3){\oval(24,8)}
\put(5,3){\circle*{4}}
\put(29,3){\circle*{4}}
\epi
& $12$&$\ep_4^2$&$1/32$
& $\int_k\De_{\rm ol}^{(1,1)}(k^2,m^2)^2=$
& $4.04576\times10^{-4}$
\\
3-2 &
\rule[-10pt]{0pt}{26pt}
\bpi(58,0)
\put(13,3){\circle{16}}
\put(29,3){\circle{16}}
\put(45,3){\circle{16}}
\put(21,3){\circle*{4}}
\put(37,3){\circle*{4}}
\epi
&$ 36 $&$ \ep_4^2 $&$ 1/16 $ 
& $J_1^2J_2=$
& $ 1/1024 $
\\\hline
3-3 &
\rule[-13pt]{0pt}{37pt}
\bpi(42,0)
\put(21,3){\circle*{4}}
\qbezier(21,3)(13.79,10.21)(13.79,13.79)
\qbezier(13.79,13.79)(13.79,21.)(21.,21.)
\qbezier(21,3)(28.21,10.21)(28.21,13.79)
\qbezier(28.21,13.79)(28.21,21.)(21.,21.)
\qbezier(21,3)(18.36,-6.85)(15.26,-8.64)
\qbezier(15.26,-8.64)(9.02,-12.24)(5.41,-6.)
\qbezier(21,3)(11.15,5.64)(8.05,3.85)
\qbezier(8.05,3.85)(1.81,0.24)(5.41,-6.)
\qbezier(21,3)(30.85,5.64)(33.95,3.85)
\qbezier(33.95,3.85)(40.19,0.24)(36.59,-6.)
\qbezier(21,3)(23.64,-6.85)(26.74,-8.64)
\qbezier(26.74,-8.64)(32.98,-12.24)(36.59,-6.)
\epi
&$ 15 $&$ -\ep_6 $&$ 1/8 $
& $J_1^3=$
& $ 1/512 $
\\\hline\hline
4-1 &
\rule[-14pt]{0pt}{34pt}
\bpi(34,0)
\put(17,3){\circle{24}}
\put(6.6,9){\line(1,0){20.8}}
\put(6.6,9){\line(3,-5){10.4}}
\put(27.4,9){\line(-3,-5){10.4}}
\put(6.6,9){\circle*{4}}
\put(27.4,9){\circle*{4}}
\put(17,-9){\circle*{4}}
\epi
&$ 288 $&$ -\ep_4^3 $&$3/512$
& $\int_k\De_{\rm ol}^{(1,1)}(k^2,m^2)^3=$
& $ 1.63237\times10^{-5} $
\\
4-2 &
\rule[-14pt]{0pt}{48pt}
\bpi(34,0)
\put(17,3){\circle{24}}
\put(17,3){\oval(24,8)}
\put(17,23){\circle{16}}
\put(5,3){\circle*{4}}
\put(29,3){\circle*{4}}
\put(17,15){\circle*{4}}
\epi
&$ 576 $&$ -\ep_4^3 $&$5/512$
& $\f{5}{8}J_1I_{3-1}=$
& $ 3.16075\times10^{-5} $
\\
4-3 &
\rule[-10pt]{0pt}{24pt}
\bpi(74,0)
\put(13,3){\circle{16}}
\put(29,3){\circle{16}}
\put(45,3){\circle{16}}
\put(61,3){\circle{16}}
\put(21,3){\circle*{4}}
\put(37,3){\circle*{4}}
\put(53,3){\circle*{4}}
\epi
&$ 432 $&$ -\ep_4^3 $&$1/64$
& $J_1^2J_2^2=$
& $ 1/16384 $
\\
4-4 &
\rule[-18pt]{0pt}{50pt}
\bpi(53.7,0)
\put(26.85,3){\circle{16}}
\put(26.85,19){\circle{16}}
\put(13,-5){\circle{16}}
\put(40.7,-5){\circle{16}}
\put(26.85,11){\circle*{4}}
\put(19.95,-1){\circle*{4}}
\put(33.75,-1){\circle*{4}}
\epi
&$ 288 $&$ -\ep_4^3 $&$3/128$
& $J_1^3J_3=$
& $ 3/32768 $
\\\hline
4-5 &
\rule[-14pt]{0pt}{34pt}
\bpi(50,0)
\put(17,3){\circle{24}}
\put(17,3){\oval(24,8)}
\put(5,3){\circle*{4}}
\put(29,3){\circle*{4}}
\put(37,3){\circle{16}}
\epi
&$ 360 $&$ \ep_4\ep_6 $&$1/64$
& $J_1I_{3-1}=$
& $ 5.05719\times10^{-5} $
\\
4-6 &
\rule[-16pt]{0pt}{37pt}
\bpi(58,0)
\put(21,3){\circle*{4}}
\put(29,3){\circle{16}}
\put(37,3){\circle*{4}}
\put(45,3){\circle{16}}
\qbezier(21,3)(10.8,3.)(8.27,5.53)
\qbezier(8.27,5.53)(3.17,10.63)(8.27,15.73)
\qbezier(21,3)(21.,13.2)(18.47,15.73)
\qbezier(18.47,15.73)(13.37,20.83)(8.27,15.73)
\qbezier(21,3)(21.,-7.2)(18.47,-9.73)
\qbezier(18.47,-9.73)(13.37,-14.83)(8.27,-9.73)
\qbezier(21,3)(10.8,3.)(8.27,0.47)
\qbezier(8.27,0.47)(3.17,-4.63)(8.27,-9.73)
\epi
&$ 540 $&$ \ep_4\ep_6 $&$1/32$
& $J_1^3J_2=$
& $ 1/8192 $
\\\hline
4-7 &
\rule[-16pt]{0pt}{38pt}
\bpi(42,0)
\put(21,3){\circle*{4}}
\qbezier(21,3)(21.,13.2)(23.53,15.73)
\qbezier(23.53,15.73)(28.63,20.83)(33.73,15.73)
\qbezier(21,3)(31.2,3.)(33.73,5.53)
\qbezier(33.73,5.53)(38.83,10.63)(33.73,15.73)
\qbezier(21,3)(10.8,3.)(8.27,5.53)
\qbezier(8.27,5.53)(3.17,10.63)(8.27,15.73)
\qbezier(21,3)(21.,13.2)(18.47,15.73)
\qbezier(18.47,15.73)(13.37,20.83)(8.27,15.73)
\qbezier(21,3)(21.,-7.2)(18.47,-9.73)
\qbezier(18.47,-9.73)(13.37,-14.83)(8.27,-9.73)
\qbezier(21,3)(10.8,3.)(8.27,0.47)
\qbezier(8.27,0.47)(3.17,-4.63)(8.27,-9.73)
\qbezier(21,3)(31.2,3.)(33.73,0.47)
\qbezier(33.73,0.47)(38.83,-4.63)(33.73,-9.73)
\qbezier(21,3)(21.,-7.2)(23.53,-9.73)
\qbezier(23.53,-9.73)(28.63,-14.83)(33.73,-9.73)
\epi
&$ 105 $&$ -\ep_8 $&$1/16$
& $J_1^4=$
& $ 1/4096 $
\\\hline\hline
5-1 &
\rule[-14pt]{0pt}{34pt}
\bpi(34,0)
\put(17,3){\circle{24}}
\put(8.5,-5.5){\line(1,0){17}}
\put(8.5,-5.5){\line(0,1){17}}
\put(8.5,11.5){\line(1,0){17}}
\put(25.5,-5.5){\line(0,1){17}}
\put(8.5,-5.5){\circle*{4}}
\put(8.5,11.5){\circle*{4}}
\put(25.5,-5.5){\circle*{4}}
\put(25.5,11.5){\circle*{4}}
\epi
&$ 2592 $&$ \ep_4^4 $&$5/4096$
& $\int_k\De_{\rm ol}^{(1,1)}(k^2,m^2)^4=$
& $ 7.55133\times10^{-7} $
\\
5-2 &
\rule[-22pt]{0pt}{48pt}
\bpi(42,0)
\put(21,-9){\circle{16}}
\put(21,15){\circle{16}}
\put(13,-9){\line(1,0){16}}
\put(13,15){\line(1,0){16}}
\put(13,3){\oval(16,24)[l]}
\put(29,3){\oval(16,24)[r]}
\put(13,-9){\circle*{4}}
\put(13,15){\circle*{4}}
\put(29,-9){\circle*{4}}
\put(29,15){\circle*{4}}
\epi
&$ 2304 $&$ \ep_4^4 $&$19/12288$
& $\int_k\De(k^2,m^2)^2\De_{\rm ss}(k^2,m^2)^2=$
& $ 1.04187\times10^{-6} $
\\
5-3 &
\rule[-18pt]{0pt}{40pt}
\bpi(58,0)
\put(13,3){\circle{16}}
\put(45,3){\circle{16}}
\put(5,-5){\line(0,1){16}}
\put(25,-5){\oval(40,16)[b]}
\put(25,11){\oval(40,16)[t]}
\put(45,3){\oval(48,16)[l]}
\put(5,3){\circle*{4}}
\put(21,3){\circle*{4}}
\put(45,-5){\circle*{4}}
\put(45,11){\circle*{4}}
\epi
&$ 10368 $&$ \ep_4^4 $&$7/6144$
& $\int_k\De_{\rm ol}^{(1,1)}(k^2,m^2)\De_{\rm eye}(k^2,m^2)=$
& $6.71540\times10^{-7}$
\\
5-4 &
\rule[-14pt]{0pt}{48pt}
\bpi(34,0)
\put(17,3){\circle{24}}
\put(17,23){\circle{16}}
\put(6.6,9){\line(1,0){20.8}}
\put(6.6,9){\line(3,-5){10.4}}
\put(27.4,9){\line(-3,-5){10.4}}
\put(6.6,9){\circle*{4}}
\put(27.4,9){\circle*{4}}
\put(17,-9){\circle*{4}}
\put(17,15){\circle*{4}}
\epi
&$ 20736 $&$ \ep_4^4 $&$1/512$
& $\f{2}{3}J_1I_{4-1}=$
& $ 1.36031\times10^{-6} $
\\
5-5 &
\rule[-30pt]{0pt}{66pt}
\bpi(34,0)
\put(17,3){\circle{24}}
\put(17,-17){\circle{16}}
\put(17,23){\circle{16}}
\put(17,3){\oval(24,8)}
\put(5,3){\circle*{4}}
\put(17,-9){\circle*{4}}
\put(17,15){\circle*{4}}
\put(29,3){\circle*{4}}
\epi
&$ 10368 $&$ \ep_4^4 $&$13/4096$
& $J_1^2\int_k\De_{\rm ol}^{(1,2)}(k^2,m^2)^2=$
& $ 2.54723\times10^{-6} $
\\
5-6 &
\rule[-14pt]{0pt}{45pt}
\bpi(46,0)
\put(23,3){\circle{24}}
\put(13,20.3){\circle{16}}
\put(33,20.3){\circle{16}}
\put(23,3){\oval(24,8)}
\put(11,3){\circle*{4}}
\put(35,3){\circle*{4}}
\put(17,13.4){\circle*{4}}
\put(29,13.4){\circle*{4}}
\epi
&$ 6912 $&$ \ep_4^4 $&$31/8192$
& $J_1^2\int_k\De(k^2,m^2)^3\De_{\rm ss}(k^2,m^2)=$
& $ 3.09329\times10^{-6} $
\\
5-7 &
\rule[-14pt]{0pt}{64pt}
\bpi(34,0)
\put(17,3){\circle{24}}
\put(17,23){\circle{16}}
\put(17,39){\circle{16}}
\put(17,3){\oval(24,8)}
\put(5,3){\circle*{4}}
\put(17,15){\circle*{4}}
\put(17,31){\circle*{4}}
\put(29,3){\circle*{4}}
\epi
&$ 6912 $&$ \ep_4^4 $&$5/2048$
& $\f{5}{8}J_1J_2I_{3-1}=$
& $ 1.97547\times10^{-6} $
\\
5-8 &
\rule[-10pt]{0pt}{26pt}
\bpi(90,0)
\put(13,3){\circle{16}}
\put(29,3){\circle{16}}
\put(45,3){\circle{16}}
\put(61,3){\circle{16}}
\put(77,3){\circle{16}}
\put(21,3){\circle*{4}}
\put(37,3){\circle*{4}}
\put(53,3){\circle*{4}}
\put(69,3){\circle*{4}}
\epi
&$ 5184 $&$ \ep_4^4 $& $1/256$
& $J_1^2J_2^3=$
&$ 1/262144 $
\\
5-9 &
\rule[-23.85pt]{0pt}{52pt}
\bpi(66,0)
\put(13,3){\circle{16}}
\put(29,3){\circle{16}}
\put(45,3){\circle{16}}
\put(53,-10.85){\circle{16}}
\put(53,16.85){\circle{16}}
\put(21,3){\circle*{4}}
\put(37,3){\circle*{4}}
\put(49,-3.9){\circle*{4}}
\put(49,9.9){\circle*{4}}
\epi
&$ 10368 $&$ \ep_4^4 $&$3/512$
& $J_1^3J_2J_3=$
& $ 3/524288 $
\\
5-10 &
\rule[-26pt]{0pt}{56pt}
\bpi(58,0)
\put(13,3){\circle{16}}
\put(29,-13){\circle{16}}
\put(29,3){\circle{16}}
\put(29,19){\circle{16}}
\put(45,3){\circle{16}}
\put(21,3){\circle*{4}}
\put(29,-5){\circle*{4}}
\put(29,11){\circle*{4}}
\put(37,3){\circle*{4}}
\epi
&$ 2592 $&$ \ep_4^4 $&$5/512$
& $J_1^4J_4=$
& $ 5/524288 $
\\\hline
5-11 &
\rule[-18pt]{0pt}{41pt}
\bpi(42,0)
\put(21,3){\circle{32}}
\put(5,-13){\oval(32,32)[rt]}
\put(37,-13){\oval(32,32)[lt]}
\put(21,-13){\line(-1,1){16}}
\put(21,-13){\line(1,1){16}}
\put(5,3){\circle*{4}}
\put(37,3){\circle*{4}}
\put(21,-13){\circle*{4}}
\epi
&$ 5760 $&$ -\ep_4^2\ep_6 $&$7/3072$
& $\int_k\De(k^2,m^2)\De_{\rm ss}(k^2,m^2)^2=$
& $ 1.50770\times10^{-6} $
\\
5-12 &
\rule[-14pt]{0pt}{48pt}
\bpi(34,0)
\put(17,3){\circle{24}}
\put(17,23){\circle{16}}
\put(6.6,-3){\line(1,0){20.8}}
\put(6.6,-3){\line(3,5){10.4}}
\put(27.4,-3){\line(-3,5){10.4}}
\put(6.6,-3){\circle*{4}}
\put(27.4,-3){\circle*{4}}
\put(17,15){\circle*{4}}
\epi
&$ 12960 $&$ -\ep_4^2\ep_6 $&$3/1024$
& $J_1I_{4-1}=$
& $ 2.04047\times10^{-6} $
\\
5-13 &
\rule[-14pt]{0pt}{46pt}
\bpi(42,0)
\put(21,3){\circle{24}}
\put(21,3){\oval(24,8)}
\put(9,3){\circle*{4}}
\put(33,3){\circle*{4}}
\put(21,15){\circle*{4}}
\qbezier(21,15)(21.,25.2)(23.53,27.73)
\qbezier(23.53,27.73)(28.63,32.83)(33.73,27.73)
\qbezier(21,15)(31.2,15.)(33.73,17.53)
\qbezier(33.73,17.53)(38.83,22.63)(33.73,27.73)
\qbezier(21,15)(10.8,15.)(8.27,17.53)
\qbezier(8.27,17.53)(3.17,22.63)(8.27,27.73)
\qbezier(21,15)(21.,25.2)(18.47,27.73)
\qbezier(18.47,27.73)(13.37,32.83)(8.27,27.73)
\epi
&$ 4320 $&$ -\ep_4^2\ep_6 $&$5/1024$
& $\f{5}{8}J_1^2I_{3-1}=$
& $ 3.95093\times10^{-6} $
\\
5-14 &
\rule[-14pt]{0pt}{48pt}
\bpi(50,0)
\put(17,3){\circle{24}}
\put(17,3){\oval(24,8)}
\put(17,23){\circle{16}}
\put(5,3){\circle*{4}}
\put(29,3){\circle*{4}}
\put(17,15){\circle*{4}}
\put(37,3){\circle{16}}
\epi
&$ 17280 $&$ -\ep_4^2\ep_6 $&$5/1024$
& $\f{5}{8}J_1^2I_{3-1}=$
& $ 3.95093\times10^{-6} $
\\
5-15 &
\rule[-14pt]{0pt}{32pt}
\bpi(66,0)
\put(17,3){\circle{24}}
\put(17,3){\oval(24,8)}
\put(5,3){\circle*{4}}
\put(29,3){\circle*{4}}
\put(37,3){\circle{16}}
\put(45,3){\circle*{4}}
\put(53,3){\circle{16}}
\epi
&$ 4320 $&$ -\ep_4^2\ep_6 $&$1/256$
& $J_1J_2I_{3-1}=$
& $ 3.16075\times10^{-6} $
\\
5-16 &
\rule[-23.85pt]{0pt}{52pt}
\bpi(50,0)
\put(29,3){\circle{16}}
\put(37,-10.85){\circle{16}}
\put(37,16.85){\circle{16}}
\put(21,3){\circle*{4}}
\put(33,-3.9){\circle*{4}}
\put(33,9.9){\circle*{4}}
\qbezier(21,3)(10.8,3.)(8.27,5.53)
\qbezier(8.27,5.53)(3.17,10.63)(8.27,15.73)
\qbezier(21,3)(21.,13.2)(18.47,15.73)
\qbezier(18.47,15.73)(13.37,20.83)(8.27,15.73)
\qbezier(21,3)(21.,-7.2)(18.47,-9.73)
\qbezier(18.47,-9.73)(13.37,-14.83)(8.27,-9.73)
\qbezier(21,3)(10.8,3.)(8.27,0.47)
\qbezier(8.27,0.47)(3.17,-4.63)(8.27,-9.73)
\epi
&$ 6480 $&$ -\ep_4^2\ep_6 $&$3/256$
& $J_1^4J_3=$
& $ 3/262144 $
\\
5-17 &
\rule[-18pt]{0pt}{40pt}
\bpi(74,0)
\put(29,3){\circle{16}}
\put(21,3){\circle*{4}}
\put(45,3){\circle{16}}
\put(37,3){\circle*{4}}
\put(61,3){\circle{16}}
\put(53,3){\circle*{4}}
\qbezier(21,3)(10.8,3.)(8.27,5.53)
\qbezier(8.27,5.53)(3.17,10.63)(8.27,15.73)
\qbezier(21,3)(21.,13.2)(18.47,15.73)
\qbezier(18.47,15.73)(13.37,20.83)(8.27,15.73)
\qbezier(21,3)(21.,-7.2)(18.47,-9.73)
\qbezier(18.47,-9.73)(13.37,-14.83)(8.27,-9.73)
\qbezier(21,3)(10.8,3.)(8.27,0.47)
\qbezier(8.27,0.47)(3.17,-4.63)(8.27,-9.73)
\epi
&$ 6480 $&$ -\ep_4^2\ep_6 $&$1/128$
& $J_1^3J_2^2=$
& $ 1/131072 $
\\
5-18 &
\rule[-12pt]{0pt}{37pt}
\bpi(78,0)
\put(21,3){\circle*{4}}
\put(39,3){\circle*{4}}
\put(57,3){\circle*{4}}
\put(13,3){\circle{16}}
\put(65,3){\circle{16}}
\qbezier(39,3)(46.21,10.21)(49.79,10.21)
\qbezier(49.79,10.21)(57.,10.21)(57.,3.)
\qbezier(39,3)(46.21,-4.21)(49.79,-4.21)
\qbezier(49.79,-4.21)(57.,-4.21)(57.,3.)
\qbezier(39,3)(31.79,10.21)(31.79,13.79)
\qbezier(31.79,13.79)(31.79,21.)(39.,21.)
\qbezier(39,3)(46.21,10.21)(46.21,13.79)
\qbezier(46.21,13.79)(46.21,21.)(39.,21.)
\qbezier(39,3)(31.79,-4.21)(28.21,-4.21)
\qbezier(28.21,-4.21)(21.,-4.21)(21.,3.)
\qbezier(39,3)(31.79,10.21)(28.21,10.21)
\qbezier(28.21,10.21)(21.,10.21)(21.,3.)
\epi
&$ 6480 $&$ -\ep_4^2\ep_6 $&$1/128$
& $J_1^3J_2^2=$
& $ 1/131072 $
\\\hline
5-19 &
\rule[-22pt]{0pt}{49pt}
\bpi(50,0)
\put(25,3){\circle{40}}
\put(25,3){\oval(40,24)}
\put(25,3){\oval(40,8)}
\put(5,3){\circle*{4}}
\put(45,3){\circle*{4}}
\epi
&$ 360 $&$ \ep_6^2 $&$1/192$
& $\int_k\De_{\rm ss}(k^2,m^2)^2=$
& $ 3.76084\times10^{-6} $
\\
5-20 &
\rule[-14pt]{0pt}{32pt}
\bpi(66,0)
\put(13,3){\circle{16}}
\put(33,3){\circle{24}}
\put(33,3){\oval(24,8)}
\put(21,3){\circle*{4}}
\put(45,3){\circle*{4}}
\put(53,3){\circle{16}}
\epi
&$ 2700 $&$ \ep_6^2 $&$1/128$
& $J_1^2I_{3-1}=$
& $ 6.32149\times10^{-6} $
\\
5-21 &
\rule[-17pt]{0pt}{39pt}
\bpi(58,0)
\put(29,3){\circle{16}}
\put(21,3){\circle*{4}}
\put(37,3){\circle*{4}}
\qbezier(21,3)(10.8,3.)(8.27,5.53)
\qbezier(8.27,5.53)(3.17,10.63)(8.27,15.73)
\qbezier(21,3)(21.,13.2)(18.47,15.73)
\qbezier(18.47,15.73)(13.37,20.83)(8.27,15.73)
\qbezier(21,3)(21.,-7.2)(18.47,-9.73)
\qbezier(18.47,-9.73)(13.37,-14.83)(8.27,-9.73)
\qbezier(21,3)(10.8,3.)(8.27,0.47)
\qbezier(8.27,0.47)(3.17,-4.63)(8.27,-9.73)
\qbezier(37,3)(37.,13.2)(39.53,15.73)
\qbezier(39.53,15.73)(44.63,20.83)(49.73,15.73)
\qbezier(37,3)(47.2,3.)(49.73,5.53)
\qbezier(49.73,5.53)(54.83,10.63)(49.73,15.73)
\qbezier(37,3)(47.2,3.)(49.73,0.47)
\qbezier(49.73,0.47)(54.83,-4.63)(49.73,-9.73)
\qbezier(37,3)(37.,-7.2)(39.53,-9.73)
\qbezier(39.53,-9.73)(44.63,-14.83)(49.73,-9.73)
\epi
&$ 2025 $&$ \ep_6^2 $&$1/64$
& $J_1^4J_2=$
& $ 1/65536 $
\\\hline
5-22 &
\rule[-18pt]{0pt}{40pt}
\bpi(50,0)
\put(17,3){\circle{24}}
\put(17,3){\oval(24,8)}
\put(5,3){\circle*{4}}
\put(29,3){\circle*{4}}
\qbezier(29,3)(29.,13.2)(31.53,15.73)
\qbezier(31.53,15.73)(36.63,20.83)(41.73,15.73)
\qbezier(29,3)(39.2,3.)(41.73,5.53)
\qbezier(41.73,5.53)(46.83,10.63)(41.73,15.73)
\qbezier(29,3)(39.2,3.)(41.73,0.47)
\qbezier(41.73,0.47)(46.83,-4.63)(41.73,-9.73)
\qbezier(29,3)(29.,-7.2)(31.53,-9.73)
\qbezier(31.53,-9.73)(36.63,-14.83)(41.73,-9.73)
\epi
&$ 5040 $&$ \ep_4\ep_8 $&$1/128$
& $J_1^2I_{3-1}=$
& $ 6.32149\times10^{-6} $
\\
5-23 &
\rule[-19pt]{0pt}{44pt}
\bpi(58,0)
\put(21,3){\circle*{4}}
\put(39,3){\circle*{4}}
\put(47,3){\circle{16}}
\qbezier(21,3)(28.21,10.21)(31.79,10.21)
\qbezier(31.79,10.21)(39.,10.21)(39.,3.)
\qbezier(21,3)(28.21,-4.21)(31.79,-4.21)
\qbezier(31.79,-4.21)(39.,-4.21)(39.,3.)
\qbezier(21,3)(13.79,10.21)(13.79,13.79)
\qbezier(13.79,13.79)(13.79,21.)(21.,21.)
\qbezier(21,3)(28.21,10.21)(28.21,13.79)
\qbezier(28.21,13.79)(28.21,21.)(21.,21.)
\qbezier(21,3)(13.79,-4.21)(10.21,-4.21)
\qbezier(10.21,-4.21)(3.,-4.21)(3.,3.)
\qbezier(21,3)(13.79,10.21)(10.21,10.21)
\qbezier(10.21,10.21)(3.,10.21)(3.,3.)
\qbezier(21,3)(28.21,-4.21)(28.21,-7.79)
\qbezier(28.21,-7.79)(28.21,-15.)(21.,-15.)
\qbezier(21,3)(13.79,-4.21)(13.79,-7.79)
\qbezier(13.79,-7.79)(13.79,-15.)(21.,-15.)
\epi
&$ 5040 $&$ \ep_4\ep_8 $ & $1/64$
& $J_1^4J_2=$
&$ 1/65536 $
\\\hline
5-24 &
\rule[-17pt]{0pt}{40pt}
\bpi(42,0)
\put(21,3){\circle*{4}}
\qbezier(21,3)(15.6,10.43)(15.6,13.6)
\qbezier(15.6,13.6)(15.6,19.)(21.,19.)
\qbezier(21,3)(26.4,10.43)(26.4,13.6)
\qbezier(26.4,13.6)(26.4,19.)(21.,19.)
\qbezier(21,3)(12.26,0.16)(9.25,1.14)
\qbezier(9.25,1.14)(4.11,2.81)(5.78,7.94)
\qbezier(21,3)(15.6,10.43)(12.59,11.41)
\qbezier(12.59,11.41)(7.45,13.08)(5.78,7.94)
\qbezier(21,3)(21.,-6.19)(19.14,-8.75)
\qbezier(19.14,-8.75)(15.96,-13.12)(11.6,-9.94)
\qbezier(21,3)(12.26,0.16)(10.4,-2.4)
\qbezier(10.4,-2.4)(7.23,-6.77)(11.6,-9.94)
\qbezier(21,3)(29.74,0.16)(31.6,-2.4)
\qbezier(31.6,-2.4)(34.77,-6.77)(30.4,-9.94)
\qbezier(21,3)(21.,-6.19)(22.86,-8.75)
\qbezier(22.86,-8.75)(26.04,-13.12)(30.4,-9.94)
\qbezier(21,3)(26.4,10.43)(29.41,11.41)
\qbezier(29.41,11.41)(34.55,13.08)(36.22,7.94)
\qbezier(21,3)(29.74,0.16)(32.75,1.14)
\qbezier(32.75,1.14)(37.89,2.81)(36.22,7.94)
\epi
&$ 945 $&$ -\ep_{10} $&$1/32$
& $J_1^5=$
& $ 1/32768 $
\\\hline\hline
\end{longtable}